\documentclass[a4paper,fleqn,useAMS,usenatbib]{mnras}

\pdfoutput=1

\usepackage{natbib}
\usepackage{mathptmx}

\usepackage[T1]{fontenc}
\usepackage{ae,aecompl}

\usepackage[dvips]{graphicx}
\usepackage{amsmath}
\usepackage{amsfonts}
\usepackage{amssymb}
\usepackage{comment}
\usepackage{bm} 
\usepackage[dvipsnames]{xcolor}
\usepackage[]{hyperref}
\hypersetup{
	colorlinks=true,	
	urlcolor=MidnightBlue,		
	hypertexnames=true,
	plainpages=false,
	naturalnames=true,
	pdftitle={Estimation of singly-transiting K2 planet periods with Gaia},    
	pdfauthor={Sandford et al.},     
	linkcolor=WildStrawberry,          
	citecolor=ForestGreen,        
}

\hypersetup{draft}

\usepackage{amsmath}
\usepackage{amsfonts}
\usepackage{amssymb}
\usepackage{wasysym}
\usepackage{rotating}
\usepackage{longtable}
\usepackage{float}
\usepackage{booktabs}
\usepackage{verbatim}
\usepackage{array}

\newcommand{\rstar}{\ensuremath{{\rm R}_{\star}\,}}
\newcommand{\mstar}{\ensuremath{{\rm M}_{\star}\,}}
\newcommand{\teff}{\ensuremath{T_{\rm eff}\,}}

\title[Periods of K2 single transiters]{
Estimation of singly-transiting K2 planet periods with Gaia parallaxes
}
\author[Sandford et al.]{Emily Sandford$^{1}$\thanks{E-mail:
\href{mailto:esandford@astro.columbia.edu}{esandford@astro.columbia.edu}}, N\'{e}stor Espinoza$^{2}$\thanks{Bernoulli Fellow}\thanks{IAU-Gruber Fellow}, Rafael Brahm$^{3,4,5}$, and Andr\'{e}s Jord\'{a}n$^{6,5}$\\
$^{1}$Department of Astronomy, Columbia University, 550 W 120th Street, New York NY 10027, USA\\
$^{2}$Max-Planck-Institut f\"{u}r Astronomie, K\"{o}nigstuhl 17, D-69117, Heidelberg, Germany\\
$^{3}$Centre of Astro-Engineering UC, Pontificia Universidad Cat\'{o}lica de Chile, Av. Vicu\~{n}a Mackenna 4860, 782-0436 Macul, Santiago, \\
Chile\\
$^{4}$Instituto de Astrof\'{i}sica, Pontificia Universidad Cat\'{o}lica de Chile, Av. Vicu\~{n}a Mackenna 4860, 782-0436 Macul, Santiago, Chile\\
$^{5}$Millennium Institute for Astrophysics, Av.\ Vicu\~{n}a Mackenna 4860, 782-0436 Macul, Santiago, Chile\\
$^{6}$Facultad de Ingenier\'ia y Ciencias, Universidad Adolfo Ib\'a\~nez, Av.\ Diagonal las Torres 2640, Pe\~nalol\'en, Santiago, Chile
}

\date{Accepted . Received ; in original form }

\pubyear{2019}

\begin{document}
\label{firstpage}
\pagerange{\pageref{firstpage}--\pageref{lastpage}}
\maketitle

\begin{abstract}

When a planet is only observed to transit once, direct measurement of its period is impossible. It is possible, however, to constrain the periods of single transiters, and this is desirable as they are likely to represent the cold and far extremes of the planet population observed by any particular survey. Improving the accuracy with which the period of single transiters can be constrained is therefore critical to enhance the long-period planet yield of surveys. Here, we combine Gaia parallaxes with stellar models and broad-band photometry to estimate the stellar densities of K2 planet host stars, then use that stellar density information to model individual planet transits and infer the posterior period distribution. We show that the densities we infer are reliable by comparing with densities derived through asteroseismology, and apply our method to 27 validation planets of known (directly measured) period, treating each transit as if it were the only one, as well as to 12 true single transiters. When we treat eccentricity as a free parameter, we achieve a fractional period uncertainty over the true single transits of $94^{+87}_{-58}\%$, and when we fix $e=0$, we achieve fractional period uncertainty $15^{+30}_{-6}\%$, a roughly threefold improvement over typical period uncertainties of previous studies.

\end{abstract}


\section{Introduction}

Choosing an observational baseline for a transit survey places fundamental limits on which planets will be observed to transit. For a baseline $B$, a planet with orbital period $P > B$ will transit \textit{at most} once during the baseline, with the transit probability falling off as $P^{-1}$ for longer periods \citep{yee08}.

As a result, transit observations of relatively long-period exoplanets are rare, even as long-period exoplanets themselves merit intense study. Planets with long periods relative to the \textit{Kepler} baseline ($\sim 1500$ days), for example, are interesting as analogs to the outer planets of the Solar System, and as examples of planets at or beyond the snow line (e.g. \citealt{kipping16}). The shortest TESS baseline, 27.4 days for most of the sky \citep{ricker15}, relegates even habitable-zone planets around FGKM stars to the ``long-period'' regime. 

However, even when a long-period transiting exoplanet is observed, constraining that planet's orbital period (and hence its distance from its host star) is difficult unless we see at least two successive transits. The observational baseline of the \textit{Kepler} survey was so long that few single-transit candidates were observed. \cite{wang15} identify 17 single-transit candidates; \cite{uehara16} enumerate a further 23 (of which 14 are new discoveries, and 9 rediscoveries of single transits identified by the \textit{Kepler} transit search pipeline); \cite{foremanmackey16} identify a further 6. Altogether, this yields a catalogue of 46 single-transit events from the initial \textit{Kepler} mission. 

In the repurposed K2 mission, which has a shorter observational baseline over each observed field ($\sim 75$ days), \cite{osborn16} identify 7 single-transit candidates. \cite{lacourse18} catalogue 164 single-transit events, although they caution that most are likely eclipsing binaries.

\cite{villanueva18}, in contrast, estimate that TESS will observe more than $200$ single transits among its postage-stamp targets (observed at 2-minute cadence), and a further $\sim 1000$ in full-frame images (observed at 30-minute cadence). They further estimate that, if they are confirmed as planets, these single-transits will double the postage-stamp targets' yield of planets with $P > 25$ days, and increase the yield of $P > 250$ day planets tenfold. \cite{huang18}, meanwhile, predict a more modest $75$ single transits among the postage-stamp targets, and $\sim 700$ in the full-frame images, but add that the single transiters will increase TESS's yield of temperate planets around FGK stars roughly threefold.

Figuring out how to accurately constrain the period distribution of single-transiters is therefore critical to enhancing the long-period planet yield of the forthcoming TESS data, as knowledge of that distribution will be important for follow-up observations aiming at confirming the planetary nature of any signals detected. There is a simple approach, first suggested by \cite{yee08}, to constrain the period with observations of a single transit. From Kepler's third law, we can relate the period of a planet to the density of its host star ($\rho_*$) and its normalised semi-major axis $(a/R_*)$: 

\begin{equation}
P^2 = \frac{3\pi}{G} \left(\frac{a}{R_*}\right)^3 \rho_*^{-1}
\end{equation}

We may measure the single-transiter's $a/R_*$ directly by modeling its transit shape; $\rho_*$, however, must come from an independent observation of the star.

Propagating uncertainty through that equation, we may derive: 

\begin{equation}
    \frac{\sigma_P}{P} = \frac{1}{2} \sqrt{\left(\frac{\sigma_{\rho_*}}{\rho_*}\right)^2 + \left(3\frac{ \sigma_{(a/R_*)}}{a/R_*}\right)^2}
\label{eq:Punc_prediction}
\end{equation} 

If we assume that the uncertainty in $\rho_*$ will dominate over the uncertainty in $a/R_*$, the above equation means 
that a $\sim 5\%$ uncertainty in $\rho_*$ translates to a $\sim 2.5\%$ uncertainty in $P$ for the single transiter 
($\sigma_P \simeq 9$ days for an Earth analog). In summary, a precise constraint on $\rho_*$, plus a single transit observation, may yield a precise constraint on the single transiter's period $P$.

Of course, this simple order-of-magnitude calculation omits some important details 
of the transit modeling, with which one actually obtains $a/R_*$. First, the retrieved value of this parameter, which comes 
from knowledge of the overall shape of the light curve in addition to the transit duration, 
is dominated by other factors in the transit modeling, such as 
limb-darkening and whether or not the planet's orbit is assumed to be circular. These two effects can lead to large biases in the retrieval of transit 
parameters if not accounted for 
\citep[see, e.g., ][on the impact of limb-darkening on the 
retrieval of $a/R_*$]{EJ:2015}. In addition, the long cadence of the data that missions like \textit{Kepler} and TESS provide for most 
stars puts an even stricter limit on the accuracy with which $a/R_*$ 
can be retrieved \citep{Kipping:2010}. (If we were to assume that the uncertainty in $a/R_*$ dominated in Equation~\ref{eq:Punc_prediction}, a $\sim 5\%$ uncertainty in $a/R_*$ would translate to a $\sim 7.5\%$ uncertainty in $P$ for a single transiter, or $\sim 1$ month for an Earth analog.)

Nevertheless, the above approach, of constraining a single-transiter's $P$ by constraining its host star's density, has been adopted by nearly all of the single-transit-catalogue works listed above. However, in most cases, because $\rho_*$ is not especially well-constrained by many types of stellar observations, they derive large $P$ uncertainties for their single-transiters. 
\cite{wang15}, for example, estimate $P$ for their single-transiters using host star densities interpolated from isochrones, achieving a typical fractional $P$ uncertainty of $\sim 100\%$. \cite{uehara16} adopt stellar density constraints from the \textit{Kepler} Community Follow-Up Observing Program, and furthermore assume circular orbits for their single-transiters, which limits the achievable $P$ precision severely. \cite{foremanmackey16} fit transit models, including inferred $P$, using priors on stellar mass and radius from the \textit{Kepler} DR25 Stellar Properties Catalog \citep{mathur17} and derive similarly large uncertainties on $P$ (the typical fractional uncertainty for their single transiters is also $\sim 100 \%$). \cite{osborn16} estimate stellar parameters by deriving stellar temperatures from broad-band colours, then calculating stellar mass and radius from those temperatures under the assumption that their stars were on the main sequence, and derive typical fractional uncertainty $\sim 50 \%$; however, they still assume that these single transiters are on circular orbits. 

Other methods of stellar characterisation, however, can yield significantly more precise constraints on $\rho_*$, which may translate to correspondingly narrow bounds on $P$ for single-transiters. Asteroseismology, for example, yields a typical $\rho_*$ precision of $\sim 5\%$ (see e.g. \citealt{huber13}, \citealt{silvaaguirre17}). \cite{sandford17} demonstrate that the method of ``stellar anchors,'' in which $\rho_*$ is measured by modeling the transits of a planet with independently well-constrained eccentricity, can also yield $\rho_*$ uncertainties of order $5\%$. 

In this paper, we investigate in detail the performance of a third method of constraining $\rho_*$: combining stellar radius measurements derived from Gaia DR2 parallaxes \citep{gaia18} with stellar mass measurements from isochrone fitting. We apply such stellar density constraints to long-period planets observed by K2. First, we use these stellar density constraints to model individual transits of known-period K2 planets \textit{as if they were single-transiters}, infer their periods, and investigate the precision and accuracy of the inferences. We then apply the method to 9 true single-transiters observed by K2. Throughout, we treat eccentricity as a free parameter in the transit model fits.  

\section{Method}

 \subsection{Stellar density estimation} \label{subsec:densities}
 In order to determine the stellar physical parameters, we follow a procedure similar to that presented in \citet{brahm:2018,k2-232}. For a given star, we first
 estimate the stellar radius by combining its publicly available photometry with
 its GAIA DR2 parallax  measurement. For this step we also require an estimate of
 the stellar atmospheric parameters in order to select a spectral energy distribution
 (SED) model to represent the star being analysed. For all the
 systems analysed in the present study we adopt the \texttt{BT-Settl-CIFIST}
 \citep{baraffe:2015} SED models. We consider the following
 sets of photometric surveys/bands in our analysis: APASS \citep[V,B,g,r,i;][]{apass},
 2MASS \citep[J,H,Ks;][]{2mass}, WISE \citep[W1,W2,W3;][]{wise}.
 For each star we construct the observed reddened emitted flux
 density at the surface of the star:
\begin{equation}
 \vec{F_o} = 4 \pi d_{gaia}^2 \cdot \vec{f},
\end{equation}
 where $d_{gaia}$ is the distance to the star computed from the Gaia
 parallax, $\vec{f}$ are the flux densities computed from the observed magnitudes,
 and the vectors have a length equal to the number of passband filters that are being  considered. We compute the uncertainties in $\vec{F_o}$ by propagating the uncertainties on the observed magnitudes and parallax. The distances are estimated using the estimation procedures presented in \citet{distance}.

Meanwhile, our model for the reddened emitted flux density at the surface of the star takes the form of a vector $\vec{F}$, where the $m^{\mathrm{th}}$ component is given by:
\begin{equation}
 F_m = 4 \pi R_{*}^2 f_m \cdot e^{-A_{\lambda}},
\end{equation}
where $R_{*}$ is the stellar radius, $f_m$ is the synthetic flux density in bandpass $m$
generated from the \texttt{BT-Settl-CIFIST} SED, and $A_{\lambda}$ is the reddening or extinction factor in that bandpass.

We assume that reddening follows the \citet{cardelli:89} law, and therefore we consider just a single reddening parameter $A_V$ in our model, which combined with the reddening law generates 
extinction factors for each passband filter. With the model and observed flux vectors in hand, we explore the posterior distributions for $R_{*}$ and $A_V$
using the \texttt{emcee} package \citep{emcee} and a log-likelihood given by $\log \mathcal{L} = \sum_i (F_{o,i}-F^m_i)^2/\sigma^2_{F_{o,i}}$. We adopt uniform priors in $R_{*}$ ([0.1$R_{\odot}$, 100$R_{\odot}$])and $A_V$ ([0, 1]) .

Once the stellar radius is estimated, we proceed to estimate the stellar mass and age
by comparing the radius and effective temperature given by stellar
evolutionary models to the observed values for these parameters. Specifically,
we use the Yonsei-Yale isochrones \citep{yi:2001} as our model, where we
fix the metallicity to the reported value, and we use the interpolating code
provided with the isochrones for generating a set of modeled \rstar\ and \teff from an arbitrary stellar mass and age. The distributions for \mstar and AGE${_\star}$
are explored using the \texttt{emcee} package. Again we use uniform priors for \mstar\ ([0.4  M$_{\odot}$, 4.5 M$_{\odot}$]) and AGE${_\star}$ ([0.05 Gyr, 4.5 Gyr]).
Finally, we use the obtained distributions of the of stellar masses and radii to 
determine the distribution of the stellar bulk density:
\begin{equation}
\rho_{\star}=\frac{M_{\star}}{\frac{4 \pi}{3}R_{\star}^3}.
\end{equation}

The use of stellar evolutionary models for the estimation of 
stellar densities can in principle produce systematic biases
if the models are not well calibrated. In order to test the
accuracy of our derived densities, we compare the
results obtained with our method with those obtained with
an independent and more precise technique. Specifically,
we use a sample of stars that have asteroseismic
density determinations. From the study presented in
\citet{silvaaguirre15,silvaaguirre17} of {\em Kepler} stars with densities
derived through asteroseismology, we select stars that
have Gaia DR2 parallaxes and no reported companions
closer than 4\arcsec. We also select two giant stars, K2-97 and K2-132, that host
close-in planets, which have density estimations from
K2 photometry \citep{grunblatt2016,grunblatt2017,jones:2017}.

\begin{figure}
   \includegraphics[width=\columnwidth]{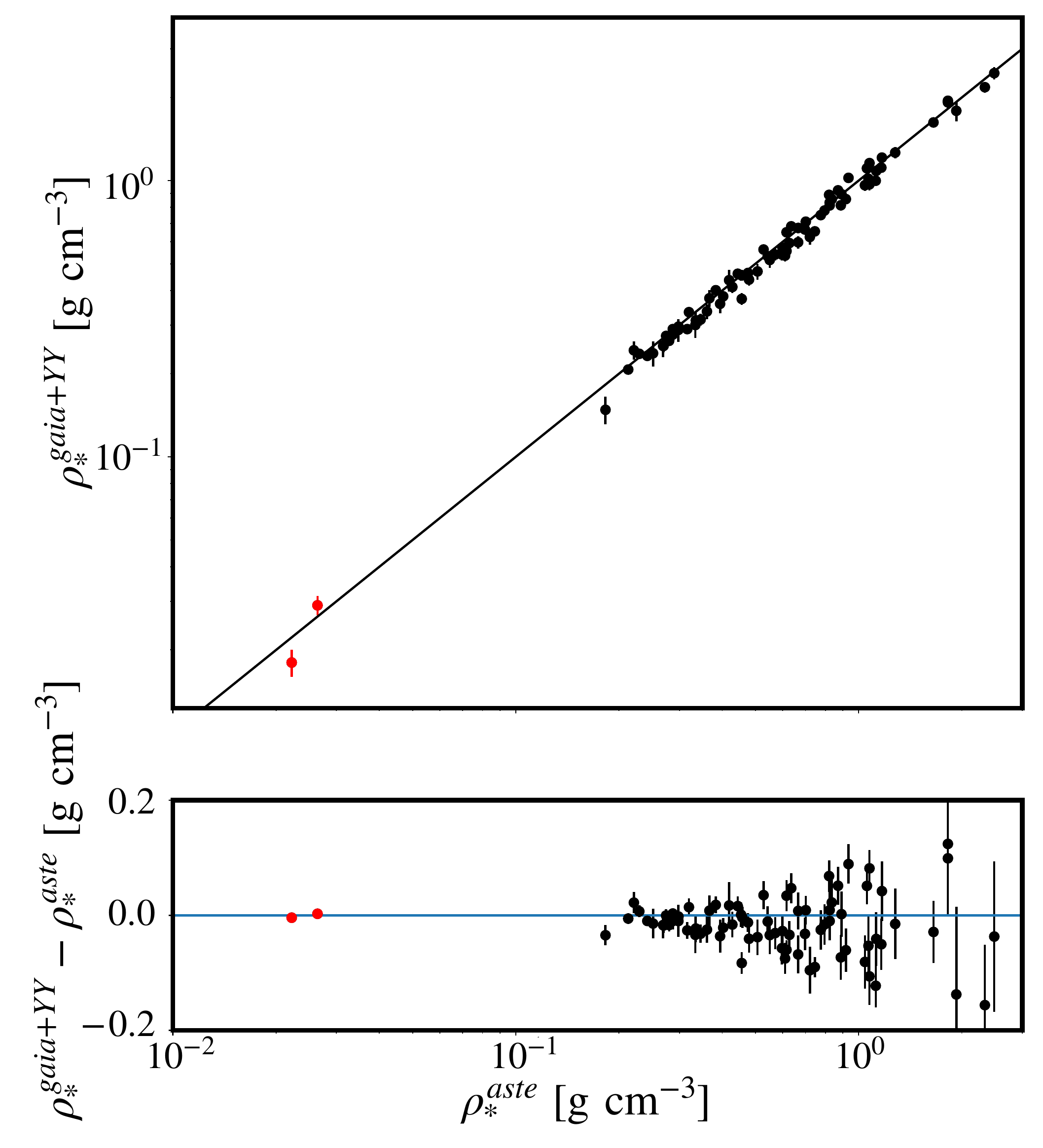}
    \caption{The top panel shows a comparison between stellar
    densities estimated with asteroseismology ($x$-axis) and those
    computed using Gaia parallaxes and the Yonsei-Yale isochrones ($y$-axis).
    The black points correspond to the sample of Kepler
    host stars presented in \citet{silvaaguirre15,silvaaguirre17}, while
    the red points correspond to the two giant stars that have been found to have
    transiting giant planets using K2 data, K2-97 and K2-132.
    The bottom panel presents the density difference between the two methods
    as a function of the asteroseismic density.}
    \label{fig:densities}
\end{figure}

We use the methods described above to compute the densities for 
these stars, adopting for them the stellar atmospheric
parameters reported in the literature. Figure~\ref{fig:densities}
compares our density estimates to those obtained through asteroseismology. The
densities range from 0.02 g cm$^{-3}$
to 2.5 g cm$^{-3}$, which is equivalent to a red giant and
a K-type dwarf, respectively. The densities computed with our method
are consistent to those obtained through asteroseismology and the residuals show no significant biases or trends between the two methods.
We also find that the residuals present a root mean square of 
0.048 g cm$^{-3}$, which is slightly larger than our mean
uncertainty in density (0.033 g cm$^{-3}$), and could signify
that our method and/or the asteroseismic method underestimates
the uncertainty in density (by $\approx 40\%$ at maximum).

To confirm that the two methods are consistent, we compute the Bayesian evidence for seven possible relationships between our stellar densities and those obtained via asteroseismology:
\begin{enumerate}
    \item A one-to-one relationship with an additional noise term,  $\rho_*^{gaia+YY} = \rho_*^{aste} + \sigma^2_{extra}$;
    \item A relationship with a constant offset, $\rho_*^{gaia+YY} = \rho_*^{aste} + b$, with and without $\sigma^2_{extra}$;
    \item A relationship with a stellar-density dependent offset, 
    $\rho_*^{gaia+YY} = a\rho_*^{aste}$, with and without $\sigma^2_{extra}$; and
    \item A relationship with both a constant and stellar-density dependent offset, $\rho_*^{gaia+YY} = a\rho_*^{aste} + b$, with and without $\sigma^2_{extra}$.
\end{enumerate}

Of these, model (i) has the highest Bayesian evidence, with measured $\sigma^2_{extra} = 0.0313 \pm 0.0044 \mathrm{g}/\mathrm{cm}^3$, which is roughly equal to our mean uncertainty in density. We therefore conclude that we have underestimated our error bars by roughly a factor of two.

We conclude from this exercise that the stellar densities
 determined using Gaia parallaxes and the Yonsei-Yale
isochrones are reliable estimates to characterise the
host stars of single transiters in order to predict their
orbital periods.

\begin{figure*}
   \includegraphics[height=0.73\columnwidth]{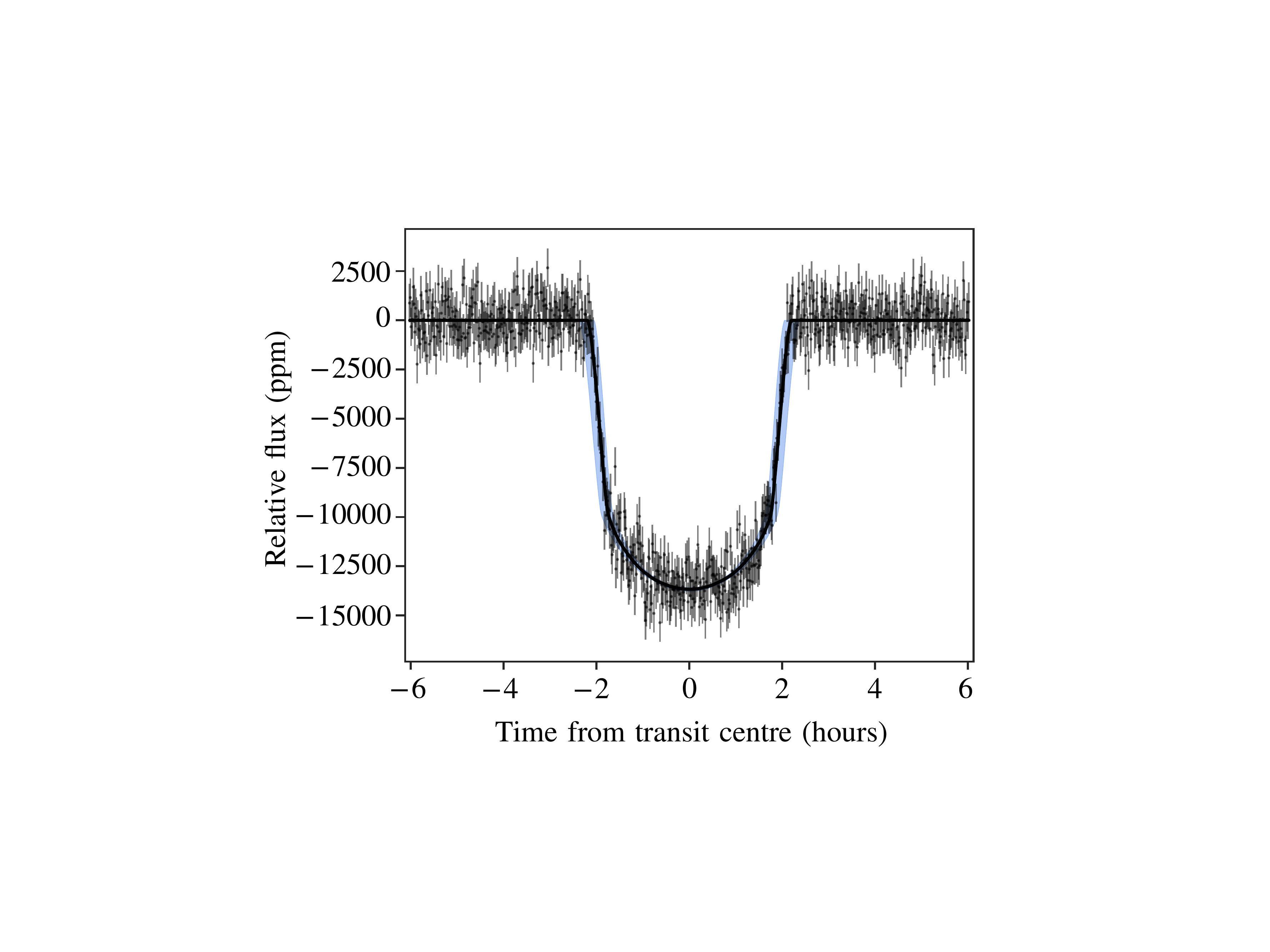}
   \includegraphics[height=0.73\columnwidth]{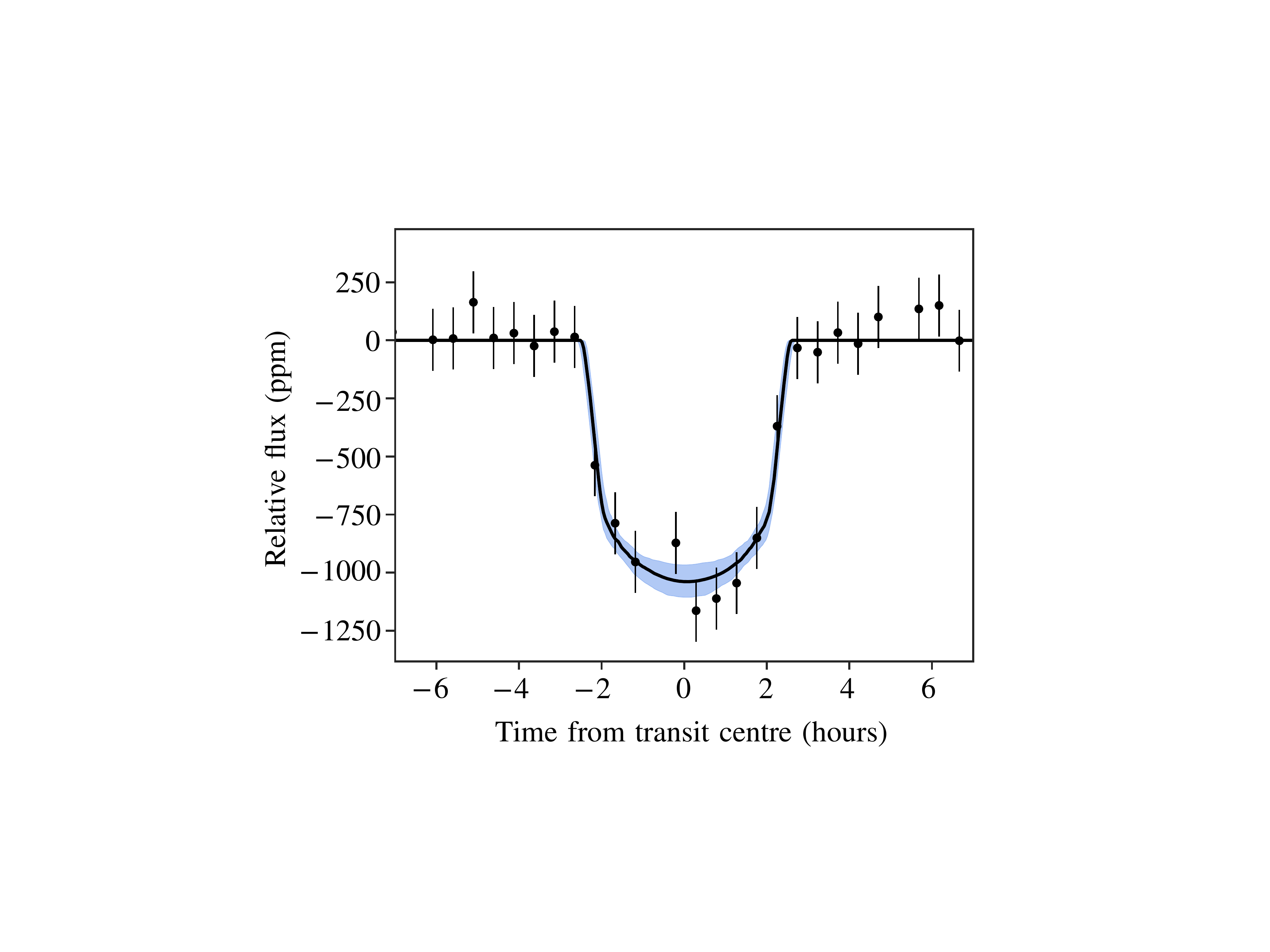}
    \caption{Illustrative plots of our single-transit fits to known exoplanets 
    HATS-11b (short cadence, left) and K2-96c (long cadence, right) 
    from K2 photometry (black dots with error bars). Solid black lines 
    present our best-fit models; blue bands the 1-sigma credibility 
    band given our posterior parameters. The out-of-transit trend, which we fit by Gaussian process regression, has been subtracted off of the K2 data.}
    \label{fig:lc_fits}
\end{figure*}

\subsection{Light curve analysis}\label{subsec:lc}

Fitting single-transit light curves is a complex problem on its own, as it 
entails sampling parameters that are strongly correlated with one another 
due to the fact that we are extracting information of several parameters from 
the same portions of the light curve \citep{SM:2003,Winn:2010}. For example, the transit duration 
simultaneously gives information on every parameter that defines the 
transit light curve except for the limb-darkening parameters, 
whereas the ingress and egress times (which are slightly 
different if we assume eccentric orbits) constrain all the 
parameters, including the limb-darkening coefficients. All this implies 
that complex, possibly multi-modal solutions are possible for a given 
transit light curve. 

Because of these possible 
complexities, we fit these transits using \texttt{MultiNest} \citep{MultiNest} through 
the \texttt{PyMultiNest} package \citep{PyMultiNest}, which allows us to 
efficiently explore the posterior distribution of the parameters 
given the data in this potentially multi-modal and degenerate case, and thus marginalise  over those possibilities when estimating the posterior distribution of the  periods of our single transiters given the data.

In our transit fits, we assume as free parameters $\vec{\theta}$ all the physical and orbital parameters that define the transit light curve: the planet-to-star radius ratio $R_p/R_*$ and the transit impact parameter $b$, reparametrised as described in \cite{espinoza18} for improved sampling efficiency; the stellar density $\rho_*$; the period $P$; the time of transit centre $t_0$; the argument of periastron passage $\omega$; and the eccentricity $e$. We describe the limb-darkening effect with a linear law through a parameter $q$, so the stellar intensity profile has the form $I(\mu) = 1 -q(1-\mu)$ with $q \in (0,1)$ in order to sample physically plausible intensity profiles. The reason for selecting a linear law is the small number of informative, in-transit data points in any given single long-cadence transit: in such a situation, the variance generated by other laws is greater than the bias generated by the linear law (see e.g. \citealt{EJ:2016}). To allow a like-for-like comparison between all of the transits in our validation sample, we apply the same linear law to the short-cadence transits as well. However, in general, single transiters should be assessed individually in order to find the best limb-darkening law for a given candidate following the prescriptions of \citet{EJ:2016}.

To account for out-of-transit trends in the light curve, we simultaneously fit (i) the eight-parameter transit model described above and (ii) a Gaussian process model with an exponential squared kernel in time, implemented in \texttt{george} \citep{dfmgeorge}. This model demands an additional four parameters: a constant flux offset $f_0$; a ``jitter" term $j$ to model the white noise in the light curve ($j$ equals the natural logarithm of the white noise variance added to the diagonal of the covariance matrix); and the two hyperparameters of the exponential squared kernel (a multiplicative constant and a scale parameter). 

We keep 30 and 100 out-of-transit points at each side of our long- and short-cadence transits, respectively, and fit the combined transit and GP model to this subset of the full time series. Our likelihood, which we assume to be Gaussian, is constructed from the light curve information, packed in a vector $\vec{y}_{tr}$ 
which contains times, fluxes and errors from the photometry. To compute the transit model, we use the \texttt{batman} package 
\citep{batman}, which is re-sampled for the case of K2 30-minute 
cadence light curves following \cite{Kipping:2010}.

For both the validation transits and the true singles, we fit the EVEREST CBV-corrected, detrended K2 light curves \citep{luger16}. We perform no outlier rejection for the validation transits; for the singles, we exclude a single $60\sigma$ discrepant out-of-transit data point from the light curve of EPIC 211311380d, a single $40\sigma$ discrepant in-transit data point from the transit of EPIC 211311380f, and a series of five $5\sigma$ discrepant in-transit data points (apparently the result of stellar activity) from the transit of EPIC 203311200b. These excluded points are plotted with the rest of the transit in Figure~\ref{fig:singles_pg2}.

We impose a split-normal prior on $\rho_*$, with mode and upper and lower standard deviations derived from the procedure described in ~\ref{subsec:densities}. We assume wide uniform priors for the reparametrised $b$ and $R_p/R_*$; $t_0$; $e$; $\omega$; $q$; and $f_0$, and wide log-uniform priors for $P$, $j$, and the two Gaussian process kernel hyperparameters. Specifically, for our validation K2 planets, we choose the $P$ prior to be a log-uniform distribution bounded between 1.0 and 1000 days.

For both our validation transits and the true single transits, we also tested the single-transit period prior suggested by \cite{kipping18} (hereafter K18), which explicitly accounts for the fact that the planet was only observed to transit once over the observational baseline, as well as the ``phase" of the single transit relative to the beginning and end of the observation. As in that work, we set $\alpha$ (the power-law index on the underlying intrinsic period prior) equal to $-2/3$, which implies uniformity over semi-major axis; the end result is a prior which is $\propto P^{-7/3}$ over the interval $P_{\mathrm{min}}$ and $P_{\mathrm{max}}$ and zero elsewhere.

For the true single transiters, we have a meaningful, observed $P_{\mathrm{min}}$, based on when the transit is observed ($t_0$) relative to the beginning ($t_{\mathrm{start}}$) and end ($t_{\mathrm{end}}$)  of its K2 campaign,

\begin{equation}
P_{\mathrm{min}} = \max [|t_0 - t_{\mathrm{start}}|, |t_0 - t_{\mathrm{end}}|].
\end{equation}
\label{eq:Pmin}

However, for the validation sample, in which our observational baseline is in fact longer than the planet's period, we must choose $P_{\mathrm{min}}$ arbitrarily. When we adopt a K18 prior with arbitrary $P_{\mathrm{min}}$ (specifically, $P_{\mathrm{min}} = 1.0$ days to enable comparison to the log-uniform prior described above), we find that the validation fits universally converge to solutions with $P = P_{\mathrm{min}} = 1.0$ days, regardless of their true periods. In these fits, the eccentricity converges to implausibly high values in order to maintain the observed transit duration in the face of such a short $P$---this is true regardless of whether the prior on $e$ is uniform, or instead a Beta distribution with parameters adopted from  \cite{kipping13}. We conclude from this exercise that, while the K18 prior is appropriate and philosophically motivated for true single transiters, it is inappropriate for our artificial validation ``single" transits with arbitrary minimum period. We therefore use the log-uniform prior described above for these validation fits.

Additionally, we impose the physical constraints that $b \leq (1 + R_p/R_*)$ (this is already imposed by the $b$ and $R_p/R_*$ reparametrisation of \citealt{espinoza18}); $b \leq (a/R_*)$; and $0^{\circ} \leq i \leq 90^{\circ}$. 

Our single-transit fitter code, 
\texttt{single}, is available on 
GitHub\footnote{\url{http://www.github.com/nespinoza/single}}.Typical fits for our targets are 
presented in Figure \ref{fig:lc_fits}.

\begin{table*}
\begin{center}
\caption{The K2 validation planets. The 21 planets above the horizontal line were observed at long (30 minute) cadence, and the 6 planets below the line were observed at short (1 minute) cadence. To the left of the vertical line are known or assumed parameters; to the right are parameters we fit to each individual transit with MultiNest. Because the $t_0$ posterior distributions are generally quite symmetrical, the reported fit $t_0$ is the 50th percentile of the posterior distribution, with uncertainties given by the $16^{th}$ and $84^{th}$ percentiles. The posterior distributions of the other parameters are asymmetrical, so we fit each with a split-normal distribution and report $\mu^{+\sigma_{\mathrm{right}}}_{-\sigma_{\mathrm{left}}}$. (Note that the parameters reported in this table are summary statistics over multiple single-transit fits for each of these planets, and therefore are not representative of any of our individual transit fits. For posterior distributions of each individual single-transit fit, please contact the authors.)}\label{tab:validation}

\footnotesize
\begin{tabular}{llll|lllllll}

K2 name & EPIC ID & $P_{\mathrm{known}}$ [d] & $\rho_*\ [\mathrm{kg}/\mathrm{m}^3]$ & $P_{\mathrm{fit}}$ [d] & $t_0$ [BJD-2454833] & $R_p/R_*$ & $b$ & $\omega\ [^\circ]$ & $e$ & $q$\\
\hline
\hline 
K2-03b & 201367065 & 10.054 & $5609.8^{+153.8}_{-156.2}$ & $8.3^{+30.0}_{-2.0}$ & $1980.419^{+0.003}_{-0.002}$ & $0.035^{+0.002}_{-0.002}$ & $0.4^{+0.2}_{-0.2}$ & $39.9^{+122.2}_{-9.9}$ & $0.0^{+0.3}_{-0.0}$ & $0.7^{+0.2}_{-0.4}$\\
K2-03c & 201367065 & 24.649 & $5609.8^{+153.8}_{-156.2}$ & $4.2^{+80.0}_{-0.6}$ & $1979.270^{+0.006}_{-0.003}$ & $0.027^{+0.002}_{-0.002}$ & $0.4^{+0.3}_{-0.2}$ & $157.1^{+106.8}_{-105.6}$ & $0.0^{+0.4}_{-0.0}$ & $0.6^{+0.2}_{-0.3}$\\
K2-10b & 201577035 & 19.304 & $1698.0^{+62.1}_{-65.2}$ & $8.3^{+20.0}_{-1.0}$ & $1986.583^{+0.004}_{-0.001}$ & $0.038^{+0.0008}_{-0.001}$ & $0.4^{+0.2}_{-0.2}$ & $44.9^{+96.8}_{-14.3}$ & $0.0^{+0.2}_{-0.0}$ & $0.3^{+0.2}_{-0.1}$\\
K2-19b & 201505350 & 7.919 & $2327.0^{+62.8}_{-64.4}$ & $7.2^{+10.0}_{-0.6}$ & $1980.384^{+0.003}_{-0.003}$ & $0.078^{+0.001}_{-0.003}$ & $0.1^{+0.4}_{-0.0}$ & $85.0^{+61.5}_{-29.1}$ & $0.0^{+0.2}_{-0.0}$ & $0.3^{+0.3}_{-0.1}$\\
K2-19c & 201505350 & 11.907 & $2327.0^{+62.8}_{-64.4}$ & $7.7^{+30.0}_{-0.9}$ & $1984.274^{+0.005}_{-0.003}$ & $0.046^{+0.0009}_{-0.002}$ & $0.4^{+0.2}_{-0.2}$ & $32.0^{+118.6}_{-5.8}$ & $0.0^{+0.2}_{-0.0}$ & $0.4^{+0.4}_{-0.2}$\\
K2-32b & 205071984 & 8.992 & $2094.0^{+85.2}_{-78.8}$ & $4.3^{+20.0}_{-0.7}$ & $2067.927^{+0.001}_{-0.001}$ & $0.057^{+0.002}_{-0.002}$ & $0.2^{+0.3}_{-0.1}$ & $33.6^{+119.4}_{-5.9}$ & $0.0^{+0.3}_{-0.0}$ & $0.3^{+0.3}_{-0.1}$\\
K2-32d & 205071984 & 31.719 & $2094.0^{+85.2}_{-78.8}$ & $9.1^{+60.0}_{-0.7}$ & $2070.787^{+0.005}_{-0.006}$ & $0.038^{+0.001}_{-0.003}$ & $0.5^{+0.2}_{-0.3}$ & $47.6^{+153.6}_{-15.8}$ & $0.0^{+0.3}_{-0.0}$ & $0.3^{+0.4}_{-0.3}$\\
K2-56b & 210848071 & 41.686 & $1029.1^{+41.8}_{-38.6}$ & $5.4^{+50.0}_{-0.6}$ & $2235.531^{+0.002}_{-0.003}$ & $0.022^{+0.001}_{-0.0006}$ & $0.2^{+0.4}_{-0.1}$ & $39.4^{+136.2}_{-9.8}$ & $0.0^{+0.3}_{-0.0}$ & $0.4^{+0.3}_{-0.1}$\\
K2-96c & 220383386 & 29.845 & $1913.8^{+60.1}_{-55.6}$ & $10.0^{+70.0}_{-1.0}$ & $2561.979^{+0.003}_{-0.002}$ & $0.030^{+0.002}_{-0.0007}$ & $0.4^{+0.2}_{-0.2}$ & $35.1^{+140.8}_{-6.5}$ & $0.0^{+0.3}_{-0.0}$ & $0.6^{+0.1}_{-0.2}$\\
K2-98b & 211391664 & 10.137 & $417.5^{+21.7}_{-14.7}$ & $2.9^{+20.0}_{-0.3}$ & $2312.980^{+0.007}_{-0.004}$ & $0.030^{+0.002}_{-0.001}$ & $0.5^{+0.2}_{-0.3}$ & $42.7^{+135.0}_{-11.4}$ & $0.0^{+0.3}_{-0.0}$ & $0.4^{+0.3}_{-0.2}$\\
K2-99b & 212803289 & 18.249 & $123.3^{+54.8}_{-12.2}$ & $15.2^{+60.0}_{-2.0}$ & $2400.826^{+0.002}_{-0.002}$ & $0.042^{+0.001}_{-0.0008}$ & $0.7^{+0.1}_{-0.3}$ & $59.4^{+95.2}_{-18.5}$ & $0.0^{+0.3}_{-0.0}$ & $0.4^{+0.1}_{-0.1}$\\
K2-110b & 212521166 & 13.864 & $2879.6^{+84.1}_{-61.0}$ & $5.6^{+20.0}_{-1.0}$ & $2400.738^{+0.001}_{-0.001}$ & $0.033^{+0.001}_{-0.0009}$ & $0.3^{+0.2}_{-0.1}$ & $36.9^{+110.9}_{-5.2}$ & $0.0^{+0.2}_{-0.0}$ & $0.7^{+0.1}_{-0.3}$\\
K2-113b & 220504338 & 5.818 & $838.0^{+43.3}_{-45.1}$ & $1.3^{+6.0}_{-0.1}$ & $2565.702^{+0.002}_{-0.002}$ & $0.086^{+0.006}_{-0.006}$ & $0.6^{+0.2}_{-0.3}$ & $56.0^{+88.3}_{-9.8}$ & $0.0^{+0.2}_{-0.0}$ & $0.8^{+0.0}_{-0.3}$\\
K2-114b & 211418729 & 11.391 & $2196.4^{+97.0}_{-107.7}$ & $7.3^{+10.0}_{-2.0}$ & $2307.325^{+0.001}_{-0.001}$ & $0.113^{+0.005}_{-0.001}$ & $0.1^{+0.3}_{-0.1}$ & $186.4^{+98.8}_{-111.2}$ & $0.0^{+0.2}_{-0.0}$ & $0.6^{+0.1}_{-0.1}$\\
K2-115b & 211442297 & 20.273 & $1671.1^{+61.0}_{-50.7}$ & $10.9^{+10.0}_{-3.0}$ & $2324.157^{+0.000}_{-0.002}$ & $0.128^{+0.001}_{-0.004}$ & $0.7^{+0.0}_{-0.1}$ & $178.2^{+61.4}_{-113.5}$ & $0.0^{+0.2}_{-0.0}$ & $0.5^{+0.3}_{-0.0}$\\
K2-139b & 218916923 & 28.382 & $2129.9^{+57.5}_{-80.2}$ & $24.2^{+30.0}_{-0.5}$ & $2492.817^{+0.001}_{-0.001}$ & $0.096^{+0.004}_{-4e-05}$ & $0.0^{+0.4}_{-0.0}$ & $86.9^{+19.9}_{-21.3}$ & $0.0^{+0.2}_{-0.0}$ & $0.6^{+0.0}_{-0.1}$\\
K2-140b & 228735255 & 6.569 & $1368.1^{+51.6}_{-55.6}$ & $5.3^{+5.0}_{-0.5}$ & $2755.286^{+0.001}_{-0.001}$ & $0.114^{+0.001}_{-0.0009}$ & $0.2^{+0.1}_{-0.1}$ & $64.8^{+54.6}_{-9.5}$ & $0.0^{+0.1}_{-0.0}$ & $0.5^{+0.1}_{-0.0}$\\
K2-232b & 247098361 & 11.168 & $837.9^{+27.7}_{-32.1}$ & $10.7^{+6.0}_{-1.0}$ & $2992.353^{+0.001}_{-0.001}$ & $0.090^{+0.0005}_{-0.0008}$ & $0.4^{+0.0}_{-0.2}$ & $67.2^{+51.3}_{-1.7}$ & $0.0^{+0.1}_{-0.0}$ & $0.5^{+0.0}_{-0.0}$\\
K2-77b & 210363145 & 8.2 & $2599.0^{+49.9}_{-60.6}$ & $2.6^{+20.0}_{-2.0}$ & $2237.805^{+0.003}_{-0.003}$ & $0.028^{+0.002}_{-0.001}$ & $0.6^{+0.2}_{-0.4}$ & $69.0^{+171.7}_{-40.5}$ & $0.0^{+0.4}_{-0.0}$ & $0.0^{+0.5}_{-0.0}$\\
EPIC 248777106b & 248777106 & 11.814 & $304.3^{+48.6}_{-7.2}$ & $2.8^{+10.0}_{-0.4}$ & $3080.805^{+0.004}_{-0.002}$ & $0.037^{+0.002}_{-0.0005}$ & $0.3^{+0.3}_{-0.2}$ & $34.4^{+143.9}_{-5.0}$ & $0.0^{+0.2}_{-0.0}$ & $0.5^{+0.2}_{-0.1}$\\
\hline
HATS-9b & 217671466 & 1.915311 & $375.0^{+30.2}_{-21.6}$ & $1.4^{+2.0}_{-0.1}$ & $2547.702^{+0.000}_{-0.000}$ & $0.084^{+0.002}_{-0.0007}$ & $0.1^{+0.2}_{-0.1}$ & $61.6^{+61.5}_{-11.8}$ & $0.0^{+0.2}_{-0.0}$ & $0.6^{+0.0}_{-0.1}$\\
HATS-11b & 216414930 & 3.6191634 & $452.9^{+36.6}_{-35.2}$ & $2.0^{+5.0}_{-0.2}$ & $2545.419^{+0.000}_{-0.000}$ & $0.109^{+0.001}_{-0.001}$ & $0.1^{+0.2}_{-0.1}$ & $39.4^{+105.6}_{-5.0}$ & $0.0^{+0.2}_{-0.0}$ & $0.5^{+0.0}_{-0.1}$\\
WASP-47b & 206103150 & 4.1591399 & $942.7^{+76.3}_{-54.1}$ & $2.6^{+3.0}_{-0.2}$ & $931.349^{+0.001}_{-0.000}$ & $0.102^{+0.001}_{-0.0006}$ & $0.3^{+0.0}_{-0.2}$ & $171.6^{+84.4}_{-105.2}$ & $0.0^{+0.2}_{-0.0}$ & $0.6^{+0.0}_{-0.0}$\\
WASP-55b & 212300977 & 4.4656291 & $1050.0^{+82.2}_{-62.4}$ & $3.2^{+3.0}_{-0.5}$ & $1583.716^{+0.000}_{-0.000}$ & $0.125^{+0.001}_{-0.002}$ & $0.2^{+0.0}_{-0.1}$ & $50.5^{+84.6}_{-2.7}$ & $0.0^{+0.1}_{-0.0}$ & $0.5^{+0.0}_{-0.0}$\\
WASP-75b & 206154641 & 2.484193 & $797.1^{+42.8}_{-41.5}$ & $1.7^{+8.0}_{-0.1}$ & $1183.269^{+0.001}_{-0.001}$ & $0.098^{+0.07}_{-0.001}$ & $0.9^{+0.1}_{-0.1}$ & $75.3^{+56.5}_{-25.6}$ & $0.0^{+0.3}_{-0.0}$ & $0.0^{+0.4}_{-0.0}$\\
WASP-118b & 220303276 & 4.0460407 & $365.4^{+30.8}_{-23.5}$ & $4.4^{+4.0}_{-0.2}$ & $2590.045^{+0.000}_{-0.000}$ & $0.082^{+0.0008}_{-0.0007}$ & $0.3^{+0.1}_{-0.1}$ & $92.4^{+23.6}_{-29.7}$ & $0.0^{+0.2}_{-0.0}$ & $0.5^{+0.0}_{-0.1}$\\
\end{tabular}
\end{center}
\end{table*}

\begin{figure}
\begin{center}
\includegraphics[width=0.45\textwidth]{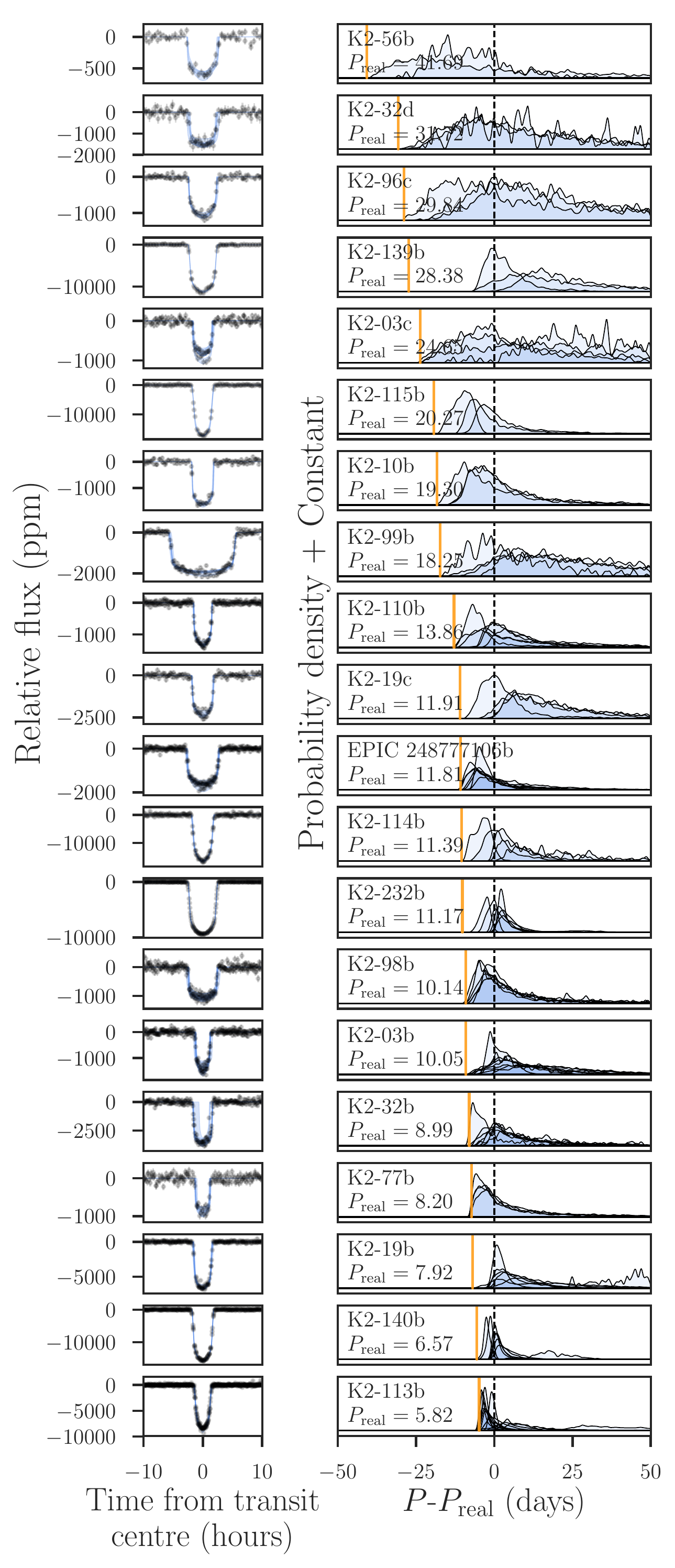}
\caption{The posterior distributions for the 21 validation planets observed at long cadence. Planets are arranged in order of increasing period from bottom to top. Each histogram represents the posterior $P$ distribution from the fit to each individual transit of the planet, with the planet's true period subtracted. The vertical yellow lines represent the lower prior bound on $P$, equal to 1.0 days. Transits with in-transit outliers and/or missing in-transit data points are susceptible to inaccurate measurement of $P$; see e.g. K2-32b and K2-140b, which both have transits with missing data points during ingress or egress.}
\label{fig:hist_period}
\end{center}
\end{figure}

\begin{figure}
\begin{center}
\includegraphics[width=0.45\textwidth]{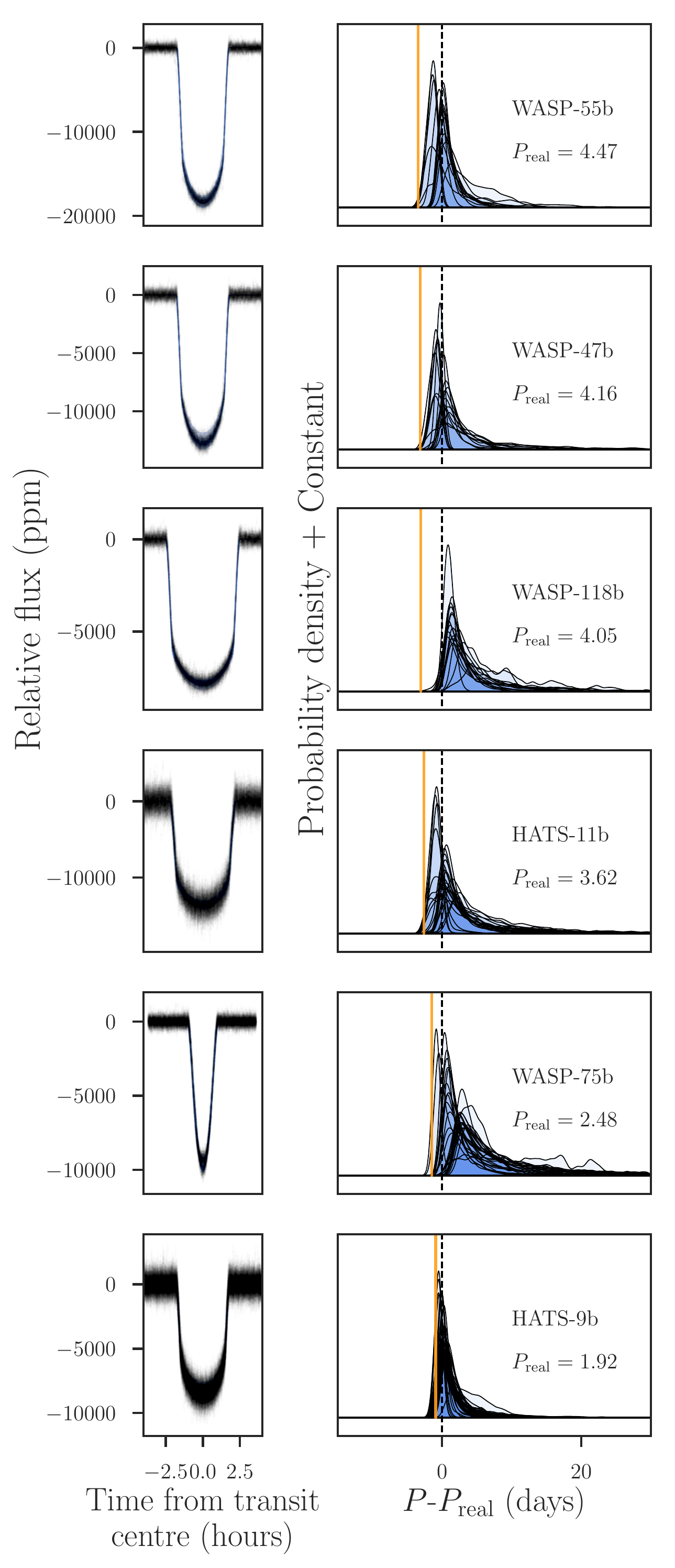}
\caption{The posterior distributions for the 6 validation planets observed at short cadence. Planets are arranged in order of increasing period from bottom to top. Each histogram represents the posterior $P$ distribution from the fit to each individual transit of the planet, with the planet's true period subtracted. The vertical yellow lines represent the lower prior bound on $P$, equal to 1.0 days.}
\label{fig:hist_period-1min}
\end{center}
\end{figure}

\begin{figure*}
\begin{center}
\includegraphics[width=\textwidth]{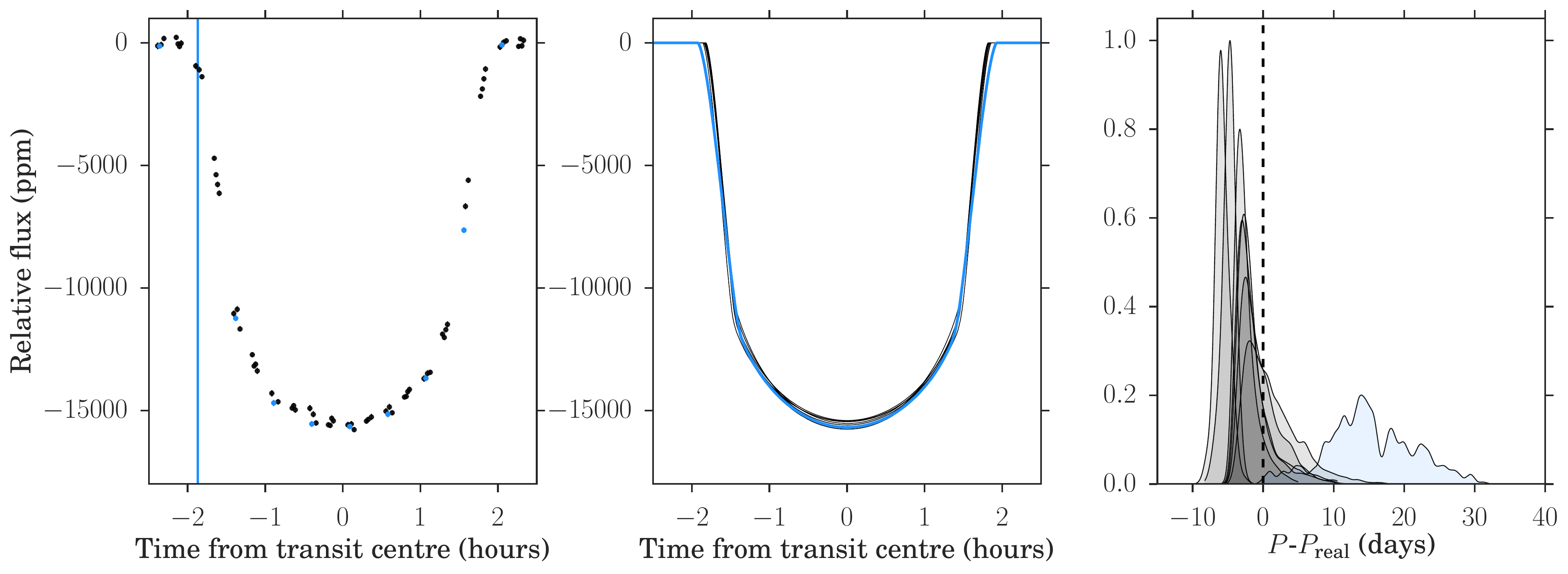}
\caption{A demonstration of the extreme effect a slightly different transit shape measurement can have on the recovered $P$ posterior, for validation planet K2-140b ($P_{real} = 6.569\ \mathrm{days}$). Eight transits are observed for this planet (left panel: transit data; middle panel: MultiNest-fit transit models; right panel: corresponding $P$ posteriors). Seven have very similar transit shapes (black data points and transit models), corresponding to very similar $P$ posteriors (black histograms). One transit, plotted in blue, has a missing data point at the position indicated by the vertical blue line in the left panel. As a result, MultiNest converges to a visibly wider transit shape, and the $P$ posterior is offset by $\sim+12.5$ days from the true $P$.}
\label{fig:shapeComparison}
\end{center}
\end{figure*}

\begin{figure}
\begin{center}
\includegraphics[width=0.48\textwidth]{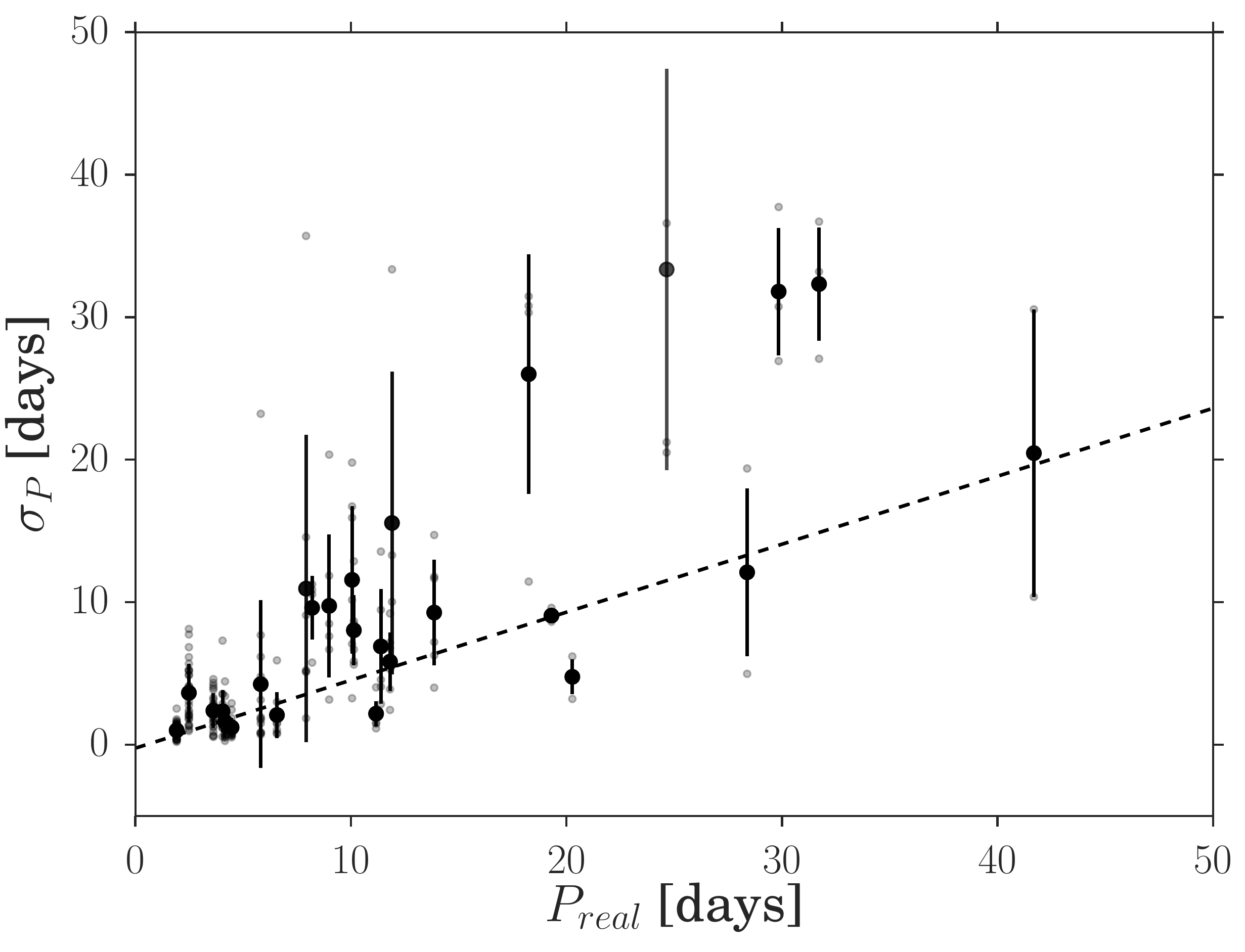}
\caption{The posterior uncertainty on $P$ as a function of $P_{\mathrm{real}}$. Semi-transparent small points represent individual transits of a given planet; larger opaque black points with error bars summarise the results over all individual transits of each planet. The dotted line is the best weighted least-squares fit. We achieve smaller uncertainties for planets with smaller orbital periods.}
\label{fig:sigmaPtrend}
\end{center}
\end{figure}

 \section{Validation Fits}
 In order to test the accuracy of our method, we applied it to known
K2 multi-transit planets, with known periods. For a given planet, we treated each transit independently as
 a single transit, and we compared our estimate of the orbital period to the
 known one. Our validation sample of 27 planets consists of the confirmed systems from K2 that have available Gaia parallaxes and stellar atmospheric parameters (which allow us in turn to obtain precise stellar masses and radii using the method outlined in Section~\ref{subsec:densities}), as well as at least one transit with SNR > 7, the threshold at which we would expect a transit to be detectable individually. 21 of the validation planets were observed at long cadence (30 minutes), but we also considered a set of 6 planets observed at 1 minute cadence.
 
 Among our validation sample were K2-19b and 19c, which underwent three overlapping transits during K2 Campaign 1. We remove these overlapping transits from our sample. We also remove one transit of WASP-47b, one transit of WASP-55b, and two transits of WASP-118b because they have in-transit light curve discontinuities due to stellar activity; the Gaussian process detrending was able to remove these discontinuities but caused the resulting de-trended transits to be unphysically much shallower than the other transits of the same planets.
 
 The systems that we analysed are listed in Table \ref{tab:validation}, along with their assumed and determined parameters. We adopted the atmospheric parameters reported
 in the discovery publications as input for the estimation of the stellar radii and masses. In the table, we report both the known period (which is measured from the interval between successive transits, and is not used in our inference in any way) and the period we measure by fitting the planet's transits individually. (Note that the parameters reported in this table are summary statistics over multiple single-transit fits for each of these planets, and therefore are not representative of any of our individual transit fits. For posterior distributions of each individual single-transit fit, please contact the authors.)
 
 Below, we discuss in turn the accuracy and precision of our period inferences for these planets.

\subsection{Accuracy of Period Inferences} \label{subsec:accuracy}

In Figures~\ref{fig:hist_period} and~\ref{fig:hist_period-1min}, we present the posterior $P$ distributions from our single-transit fits to the long- and short-cadence validation planets, respectively, as well as plots of the fits themselves. Because of the physical requirement that $P > 0$ (i.e., $P - P_{\mathrm{real}} > -P_{\mathrm{real}}$), these posterior distributions have a characteristic asymmetric shape, skewed toward long $P$. To summarise these posteriors, we therefore choose to fit each one with a split-normal distribution and report the mean and left/right standard deviations \citep{penoyre19}. (In Table~\ref{tab:validation}, the given summary statistics are from split-normal fits to the concatenated posterior samples of all individual single transit fits for a given planet.)

Generally, these posterior $P$ distributions are in good agreement with $P_{\mathrm{real}}$. This is particularly true for the six validation planets observed at short cadence, where the transit shape is much better constrained. 

Occasionally, a particular transit will yield a $P$ posterior that is significantly offset from $P_{real}$; this is never because of a poor fit to the available light curve data, as Figures~\ref{fig:hist_period} and~\ref{fig:hist_period-1min} show. Rather, these are cases where our ability to accurately measure the transit shape is compromised by missing data, outliers, or both.

In Figure~\ref{fig:shapeComparison}, we illustrate the surprisingly dramatic effect that even a small discrepancy in transit shape can have on the accuracy of the $P$ posterior, using K2-140b as an example. This planet is observed to transit eight times during K2 Campaign 10. Of our fits to the eight transits individually, seven have very similar transit shapes and return very similar $P$ posteriors, in reasonably good agreement with $P_{\mathrm{real}} = 6.569\ \mathrm{days}$. 

The seventh of eight, however, is missing a data point during ingress, and the resulting MultiNest fit has a noticeably longer transit duration than the others. The resulting $P$ posterior is centred at $\mu = 19.0^{+7.9}_{-4.0}$ days, more than $3\sigma_{\mathrm{left}}$ discrepant with $P_{\mathrm{real}}$. This is sensible because $P$ scales as $(a/R_*)^{3/2}$, so an overestimate in $a/R_*$ gets amplified in $P$. 

We note also that missing data during ingress and egress can also cause MultiNest to converge to an erroneously short transit duration, and a correspondingly short period---this is true for two of K2-32b's seven observed transits. 

We therefore advise caution in fitting single transits with missing data or obvious in-transit outliers, particularly during ingress and egress, as the fits for such transits are not reliable. The nine true single transits discussed in section~\ref{sec:predictions} are well-sampled during ingress and egress, so we have confidence in our ability to measure their transit shapes (and corresponding periods) accurately (see Figures~\ref{fig:singles_pg1} and~\ref{fig:singles_pg2}, left panels).

\begin{figure}
\begin{center}
\includegraphics[width=0.45\textwidth]{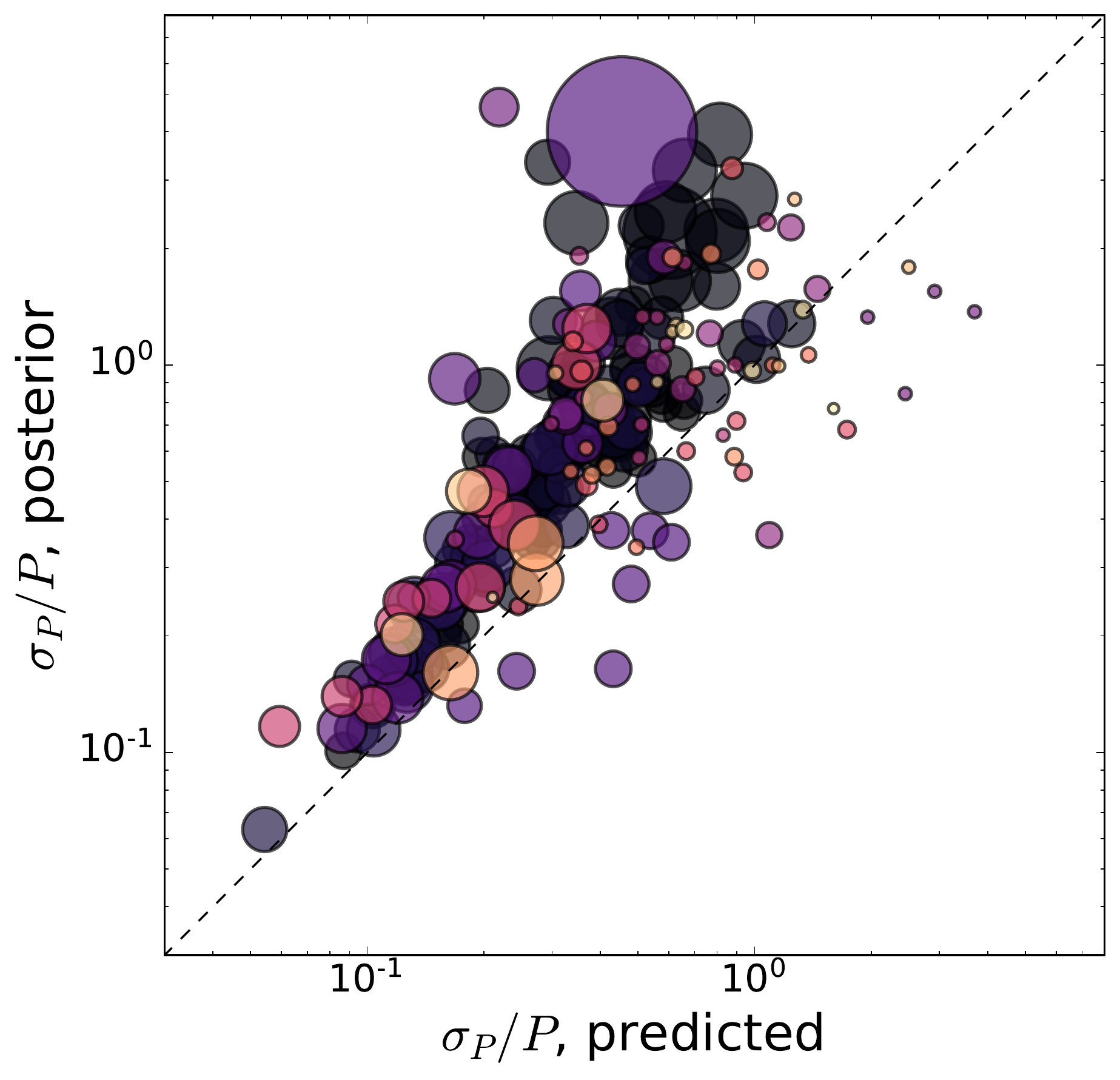}
\caption{The posterior fractional $P$ uncertainty for each single-transit fit of the validation planets, as a function of the fractional $P$ uncertainty predicted by Equation~(\ref{eq:Punc_prediction}). The black dotted line is a one-to-one line (the predicted relationship of Equation~\ref{eq:Punc_prediction}). Each data point represents one transit of one validation planet; points are colour-coded by $P_{\mathrm{real}}$ (dark for short $P_{\mathrm{real}}$ to light for large $P_{\mathrm{real}}$) and sized by the best-fit posterior $R_p/R_*$.}
\label{fig:Pcomp}
\end{center}
\end{figure}

\subsection{Precision of Period Inferences} \label{subsec:precision}

We also consider the precision of the period inferences we can make about our validation planets. In particular: are there specific characteristics of a single transit that enable a precise period measurement? 

We note that we calculate the posterior period uncertainty, $\sigma_P$, by fitting a split-normal distribution to the $P$ posterior, then taking the average of the resulting $\sigma_{\mathrm{left}}$ and $\sigma_{\mathrm{right}}$. We make this choice because our posteriors are quite asymmetric, as explained in~\ref{subsec:accuracy}, and the traditional percentile summary statistics do not characterise them well. (We adopt the same procedure for calculating posterior uncertainties $\sigma_{a/R_*}$ and $\sigma_e$, below.)

In Figure~\ref{fig:sigmaPtrend}, we examine the trend between the posterior period uncertainty $\sigma_P$ and the true period $P_{\mathrm{real}}$ for our 27 validation planets. These quantities are positively correlated: in other words, we derive smaller $\sigma_P$ for smaller $P_{\mathrm{real}}$ and vice versa. (This trend is also visible in Figure~\ref{fig:hist_period}, in the sense that the average width of the $P$ posteriors increases with increasing $P_{\mathrm{real}}$.) 


Given that $\sigma_P$ increases with $P$, we next investigate trends in the fractional period uncertainty, $\sigma_P/P$, with other quantities. First, in Figure~\ref{fig:Pcomp}, we compare the $\sigma_P/P$ from our posterior $P$ distributions to the $\sigma_P/P$ predicted by Equation~(\ref{eq:Punc_prediction}). Broadly, the data follow the expected relationship, but we find that Equation~(\ref{eq:Punc_prediction}) tends to underestimate the observed fractional uncertainty, particularly when $\sigma_P/P$ is high.

\begin{figure}
\begin{center}
\includegraphics[width=0.32\textwidth]{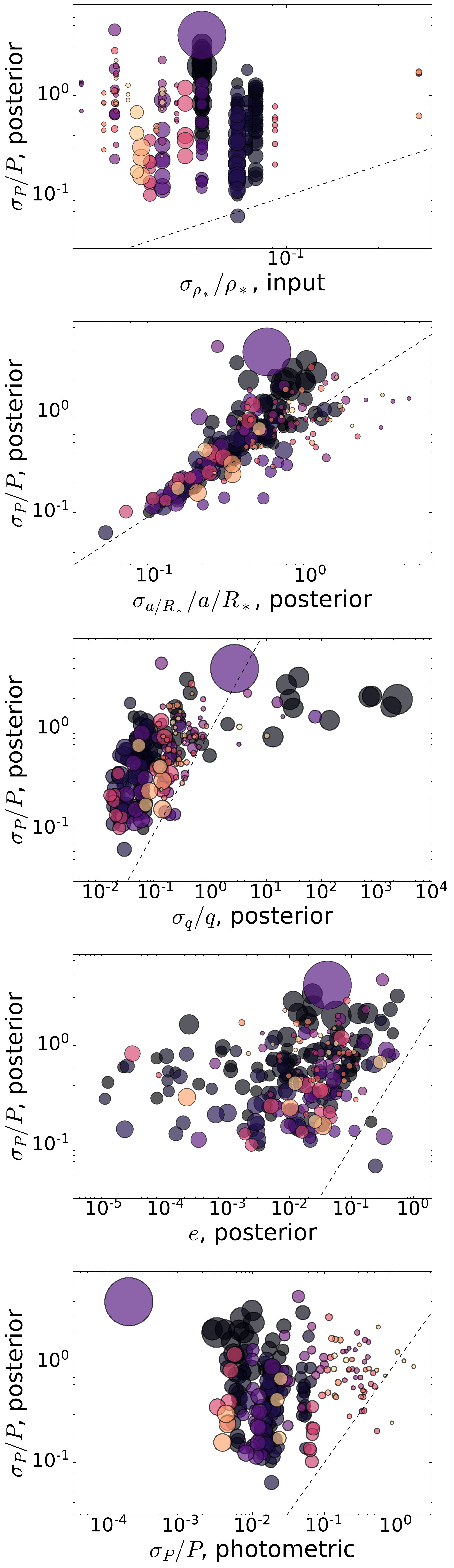}
\caption{An exploration of which terms contribute most significantly to $\sigma_P/P$. Top: $\sigma_P/P$ as a function of the fractional uncertainty on the Gaia-derived $\rho_*$ measurement input to MultiNest. Second row: $\sigma_P/P$ as a function of the fractional posterior uncertainty on normalised semi-major axis $a/R_*$. Third row: $\sigma_P/P$ as a function of the fractional posterior uncertainty on linear limb darkening coefficient $q$. Fourth row: $\sigma_P/P$ as a function of the posterior modal value of eccentricity $e$. Bottom: $\sigma_P/P$ as a function of the K2 photometric uncertainty for each transit (calculated from the formula given by \citealt{yee08}). Each point represents one transit of one validation planet; points are colour-coded by $P_{real}$ (dark for short $P_{real}$ to light for large $P_{real}$ and sized by the best-fit posterior $R_p/R_*$. A one-to-one line is plotted in each panel.}
\label{fig:Punc_explanation}
\end{center}
\end{figure}

In Figure~\ref{fig:Punc_explanation}, we investigate $\sigma_P/P$ more carefully, examining which sources of uncertainty contribute most strongly with it.  In particular, we examine the correlation of $\sigma_P/P$ with the fractional uncertainty of the input Gaia $\rho_*$ measurement; with the fractional uncertainty of $a/R_*$ and $e$ from our posterior distributions; and with the estimated contribution to $\sigma_P/P$ from K2's photometric uncertainty in observing each planet's transit. (This last quantity is calculated according to Equation 13 of \citealt{yee08}.) 

The top panel of Figure~\ref{fig:Punc_explanation} plots the relationship between the fractional uncertainty in $\rho_*$, which was a prior input to our MultiNest fits, and the resulting $\sigma_P/P$. Since $\sigma_{\rho_*}/\rho_*$ is the same for all planets in a particular system, the planets in this panel bunch up in vertical lines. Surprisingly, there is no apparent correlation between the prior fractional stellar density uncertainty and the posterior fractional uncertainty in $P$: in other words, the stellar density uncertainty does not appear to be the dominant contribution to the ultimate $P$ uncertainty. Our order-of-magnitude calculation for an Earth analog, following Equation~(\ref{eq:Punc_prediction}), is clearly too simplistic.

In contrast, the second panel shows that $\sigma_P/P$ is very strongly predicted by $\sigma_{a/R_*}/a/R_*$. $a/R_*$ is measured directly from the transit shape; this panel shows that our ability to constrain the transit shape is the single most important predictor of how well we can constrain $P$ for a single transiter. 

Since we also expect our constraint on the linear limb darkening coefficient $q$ to depend on how well we can constrain the transit shape, we plot in the third panel $\sigma_P/P$ vs. $\sigma_q/q$, and find the expected positive correlation; in other words, transits with worse-constrained shapes have both worse-constrained limb darkening coefficients and worse-constrained periods.

Our ability to constrain $e$ is less important, as illustrated in the fourth panel. There is a weak positive correlation between $\sigma_P/P$ and $e$, indicating that more eccentric planets (larger $e$) have worse period constraints, and vice versa.

Finally, there is essentially no correlation between $\sigma_P/P$ and the K2 photometric uncertainty over the planet transit. Visible in the colours and sizes of points on this plot, however, are the strong negative correlation between photometric uncertainty and planet size ($(\sigma_P/P)_{\mathrm{photometric}} \propto R_p^{-5/2}$) and the weaker negative correlation between photometric uncertainty and orbital period ($(\sigma_P/P)_{\mathrm{photometric}} \propto P^{-1/6}$).

In summary, a strong constraint on $a/R_*$, which is measured from the transit shape, is the best predictor of a strong posterior constraint on $P$. To improve the constraint on the transit shape, one could place stronger prior constraints on the limb-darkening profile of the star, for example by jointly fitting transits of multiple planets in the same system.

\begin{table*}
\begin{center}
\caption{Summary statistics for fits to the true K2 single transits. To the left of the vertical line are known or assumed parameters; to the right are parameters fit with MultiNest. Because the $t_0$ posterior distributions are generally quite symmetrical, the reported fit $t_0$ is the $50^{\mathrm{th}}$ percentile of the posterior distribution, with uncertainties given by the $16^{\mathrm{th}}$ and $84^{\mathrm{th}}$ percentiles. The posterior distributions of the other parameters are asymmetrical, so we fit each with a split-normal distribution and report $\mu^{+\sigma_{\mathrm{right}}}_{-\sigma_{\mathrm{left}}}$.}\label{tab:singles}
\footnotesize
\begin{tabular}{llll|llllllll}

Name (EPIC) & Reference & $P_{\mathrm{min}}$ [d] & $\rho_*\ [\mathrm{kg}/\mathrm{m}^3]$ & $P_{\mathrm{inferred}}$ [d] & $t_0$ [BJD-2454833] & $R_p/R_*$ & $b$ & $\omega\ [^\circ]$ & $e$ & $q$\\

\hline
\hline
EPIC 201635132b & Osborn16 & 63.3 & $4992.3^{+222.5}_{-224.7}$ & $125.9^{+42.6}_{-22.6}$ & $1993.985^{+0.001}_{-0.001}$ & $0.227^{+0.02}_{-0.02}$ & $0.9^{+0.1}_{-0.0}$ & $115.5^{+106.5}_{-58.6}$ & $0.03^{+0.09}_{-0.02}$ & $0.84^{+0.05}_{-0.32}$\\
EPIC 201892470b & Osborn16 & 78.7 & $199.8^{+15.9}_{-7.5}$ & $78.8^{+488.1}_{-0.1}$ & $2056.159^{+0.020}_{-0.030}$ & $0.026^{+0.0008}_{-0.009}$ & $1.0^{+0.0}_{-0.3}$ & $69.3^{+84.5}_{-29.5}$ & $0.88^{+0.05}_{-0.35}$ & $0.97^{+0.02}_{-0.42}$\\
EPIC 203311200b & Osborn16 & 59.7 & $325.1^{+30.2}_{-27.2}$ & $251.0^{+145.9}_{-33.9}$ & $2121.016^{+0.001}_{-0.001}$ & $0.060^{+0.0006}_{-0.0005}$ & $0.8^{+0.0}_{-0.0}$ & $88.6^{+30.4}_{-17.5}$ & $0.03^{+0.14}_{-0.02}$ & $0.44^{+0.09}_{-0.05}$\\
EPIC 204634789b & Osborn16 & 50.8 & $848.0^{+48.3}_{-53.9}$ & $64.5^{+110.4}_{-5.4}$ & $2088.016^{+0.001}_{-0.001}$ & $0.090^{+0.003}_{-0.003}$ & $0.8^{+0.0}_{-0.1}$ & $63.9^{+22.3}_{-6.4}$ & $0.92^{+0.02}_{-0.03}$ & $0.73^{+0.07}_{-0.25}$\\
EPIC 211311380d & Vanderburg16 & 48.2 & $658.0^{+26.0}_{-26.0}$ & $49.3^{+328.5}_{-0.1}$ & $2333.270^{+0.001}_{-0.001}$ & $0.026^{+0.001}_{-0.0008}$ & $0.1^{+0.3}_{-0.1}$ & $60.4^{+143.4}_{-29.8}$ & $0.02^{+0.36}_{-0.01}$ & $0.61^{+0.07}_{-0.14}$\\
EPIC 211311380e & Vanderburg16 & 72.4 & $658.0^{+26.0}_{-26.0}$ & $131.7^{+245.9}_{-12.4}$ & $2309.016^{+0.001}_{-0.001}$ & $0.036^{+0.001}_{-0.0005}$ & $0.5^{+0.1}_{-0.2}$ & $79.2^{+93.4}_{-25.8}$ & $0.02^{+0.22}_{-0.01}$ & $0.46^{+0.08}_{-0.02}$\\
EPIC 211311380f & Vanderburg16 & 47.3 & $658.0^{+26.0}_{-26.0}$ & $599.2^{+525.3}_{-159.5}$ & $2353.915^{+0.001}_{-0.001}$ & $0.066^{+0.001}_{-0.0005}$ & $0.4^{+0.1}_{-0.2}$ & $88.3^{+30.7}_{-36.6}$ & $0.01^{+0.27}_{-0.01}$ & $0.44^{+0.05}_{-0.04}$\\
EPIC 212813907b & Crossfield18 & 32.8 & $2309.8^{+188.0}_{-164.8}$ & $2014.9^{+1117.2}_{-331.6}$ & $3380.823^{+0.001}_{-0.001}$ & $0.119^{+0.001}_{-0.0003}$ & $0.2^{+0.1}_{-0.1}$ & $103.5^{+20.2}_{-3.7}$ & $0.41^{+0.08}_{-0.12}$ & $0.63^{+0.02}_{-0.02}$\\
EPIC 228801451d & Santerne18 & 25.0 & $2702.2^{+84.8}_{-104.9}$ & $25.0^{+49.8}_{-0.1}$ & $2790.062^{+0.003}_{-0.003}$ & $0.030^{+0.0003}_{-0.002}$ & $0.5^{+0.2}_{-0.2}$ & $85.0^{+37.3}_{-32.6}$ & $0.69^{+0.11}_{-0.13}$ & $0.18^{+0.25}_{-0.09}$\\
EPIC 248045685b & Vanderburg18 & 72.7 & $3082.8^{+267.4}_{-186.7}$ & $118.0^{+119.3}_{-41.4}$ & $2995.387^{+0.001}_{-0.001}$ & $0.027^{+0.0004}_{-0.0008}$ & $0.5^{+0.2}_{-0.1}$ & $95.2^{+30.3}_{-23.1}$ & $0.44^{+0.11}_{-0.16}$ & $0.26^{+0.17}_{-0.10}$\\
EPIC 248847494b & Giles18 & 59.8 & $86.6^{+8.1}_{-7.3}$ & $1658.2^{+4191.1}_{-445.8}$ & $3134.174^{+0.020}_{-0.020}$ & $0.044^{+0.002}_{-0.003}$ & $0.9^{+0.0}_{-0.1}$ & $308.2^{+19.5}_{-105.2}$ & $0.08^{+0.29}_{-0.05}$ & $0.75^{+0.08}_{-0.34}$\\
EPIC 246445793b & Vanderburg15 & 3.9 & $2750.5^{+158.8}_{-252.8}$ & $4.7^{+11.6}_{-0.2}$ & $1865.133^{+0.002}_{-0.001}$ & $0.030^{+0.002}_{-0.002}$ & $0.2^{+0.3}_{-0.1}$ & $97.1^{+37.6}_{-45.9}$ & $0.10^{+0.26}_{-0.06}$ & $0.86^{+0.09}_{-0.13}$\\

\end{tabular}
\end{center}
\end{table*}

\normalsize

\section{Period Predictions for K2 Single Transiters} \label{sec:predictions}

We next proceed to apply our fitting code to twelve single transiters observed by K2. These single transiters were first reported by \cite{osborn16, crossfield18, santerne18, vanderburg18, giles18}; \cite{vanderburg15}; and \cite{vanderburg16}, as detailed in Table~\ref{tab:singles}. 

\begin{figure*}
\begin{center}
\includegraphics[width=\textwidth]{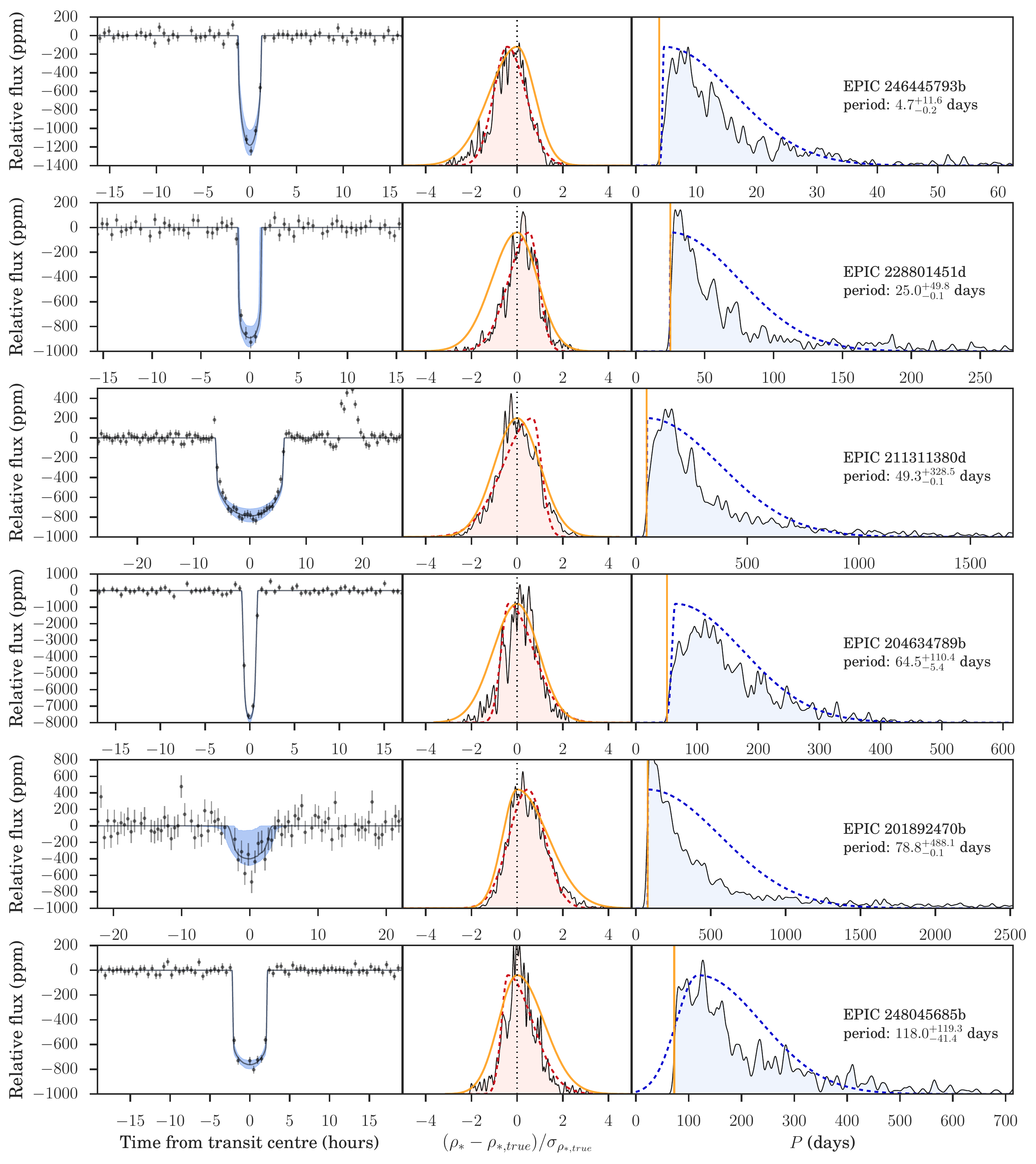}
\caption{The results of fits to the twelve true K2 single transits, compiled from sources detailed in Table~\ref{tab:singles}. Each row represents one planet. Left panel: The observed transit (black data points), best-fit model (solid black line), and 1-sigma credibility band given our posterior parameters (blue band). Middle panel: The Gaussian stellar density prior (yellow line) derived from Gaia distances plus available photometry for each host, and the corresponding posterior $\rho_*$ distribution (red histogram). Right panel: The posterior $P$ distribution, bounded on the left by the K2 baseline-deduced $P_{\mathrm{min}}$ (yellow line). We fit each $P$ posterior with a split normal distribution (blue dotted line) to allow us to write down summary statistics (right panel text).}
\label{fig:singles_pg1}
\end{center}
\end{figure*}
\begin{figure*}
\begin{center}
\includegraphics[width=\textwidth]{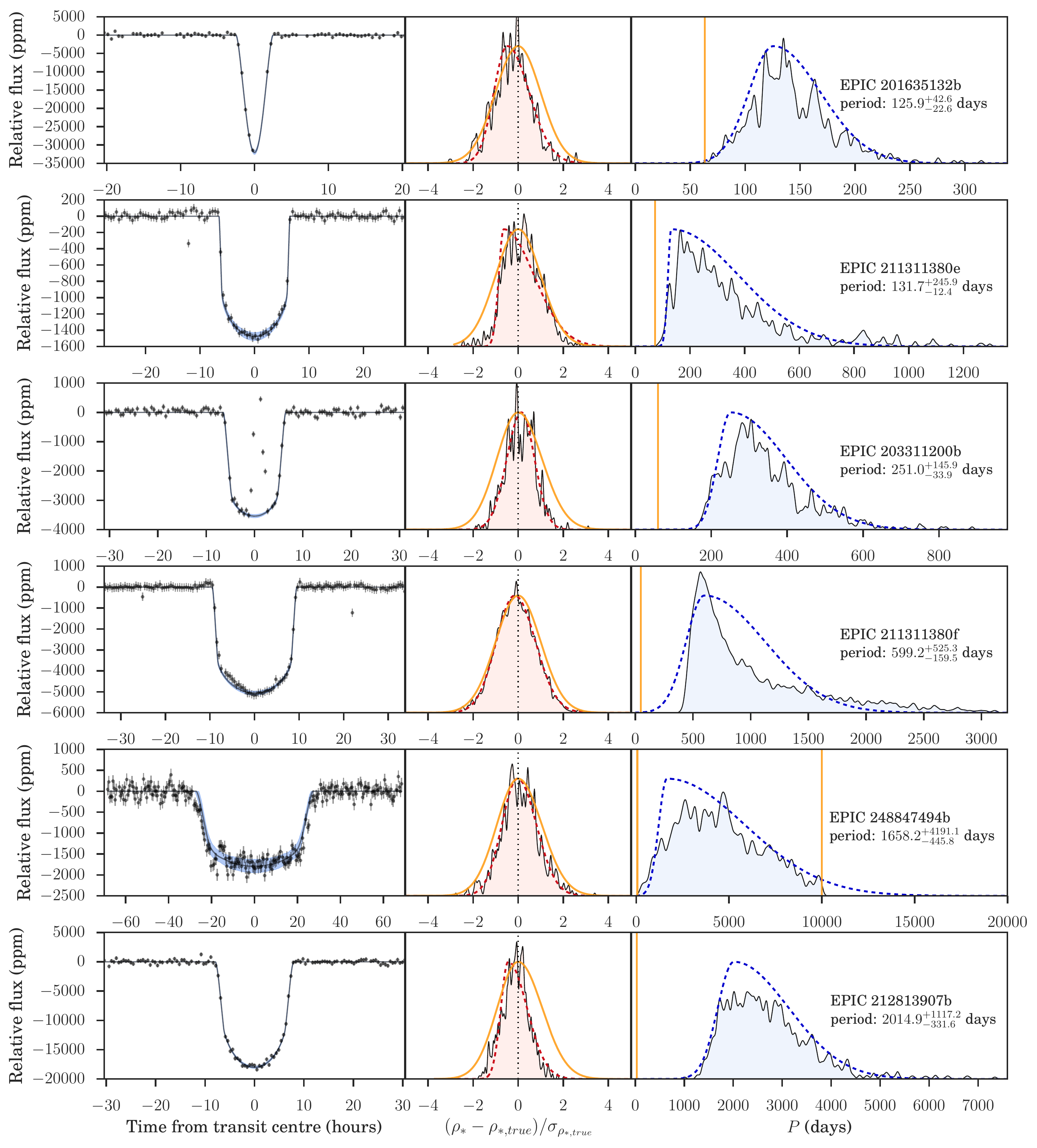}
\caption{A continuation of Figure~\ref{fig:singles_pg1}. The $P$ posterior for the second-to-last planet, EPIC 248847494b, abuts the upper end of the $P$ prior, $P_{\mathrm{max}} = 10000$ days.}
\label{fig:singles_pg2}
\end{center}
\end{figure*}

In Figures~\ref{fig:singles_pg1} and~\ref{fig:singles_pg2}, we present our transit fits and posterior distributions for these twelve single transiters. For all twelve, MultiNest converges to a good fit to the transit, with a stellar density posterior distribution in complete agreement with the Gaia prior. (In other words, there were no cases for which MultiNest needed to wander far from the input Gaia stellar density to fit the transit data.)

Over the twelve single transits, our posterior fractional uncertainty $\sigma_P/P$ is $94^{+87}_{-58}\%$, which is comparable to that achieved in previous work. We emphasise that we treat eccentricity $e$ as a free parameter in these fits.

For a more direct comparison to previous work, we re-fit the single transits with eccentricity $e$ fixed to zero, and find that we achieve posterior period fractional uncertainty $\sigma_P/P$ of $15^{+30}_{-6}\%$, a roughly threefold improvement over typical uncertainties of previous studies.

\begin{figure*}
\begin{center}
\includegraphics[width=\textwidth]{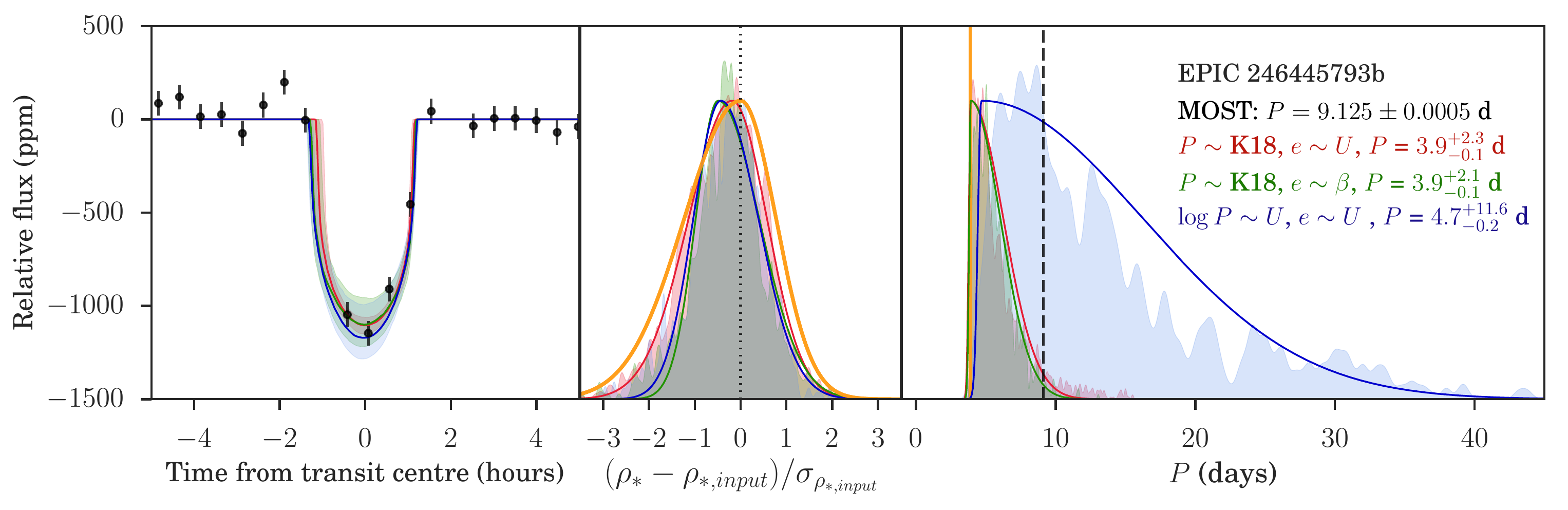}
\caption{A comparison of the period posteriors derived for single transiter EPIC 246445793b with three different choices of prior: in red, a K18 period prior with $\alpha = -2/3$ and a uniform eccentricity prior between 0 and 1; in green, a K18 period prior with $\alpha = -2/3$ and a Beta distribution eccentricity prior with $a = 0.867, b=3.03$ (adopted from \citealt{kipping13}); and in blue, a log-uniform period prior between 1 and 10000 days with a uniform eccentricity prior between 0 and 1. Left panel: The best-fit transit model given each prior. Middle panel: The stellar density posterior given each prior. Right panel: The period posterior given each prior. The dashed black line marks the MOST photometry-measured period for this planet \citep{vanderburg15}.}
\label{fig:prior_comparison}
\end{center}
\end{figure*}

\footnotetext{This single transit was observed during a 9-day test of K2 in February 2014, hence its short $P_{\min}$.}

\subsection{A note on priors}

As discussed in Section~\ref{subsec:lc}, the log-uniform prior adopted for our validation fits does not formally account for the fact that a single-transiter is only observed to transit once, and as such does not exploit all of the available information about a single transit. The prior defined by K18 does account for this fact, and it strongly enforces the expectation that, if we observe a single transit of a planet over a given baseline, the planet is a-priori less likely to have a long period than a short one ($\mathrm{Pr}(P) \propto P^{\alpha - 5/3}$ between $P_{\mathrm{min}}$ and $P_{\mathrm{max}}$, where we choose $\alpha = -2/3$, the power-law index of the underlying intrinsic period prior, for uniformity in semi-major axis).

We test both period priors on our twelve true single transiters. For both, we set $P_{\mathrm{min}}$ according to Equation~\ref{eq:Pmin} and set $P_{\mathrm{max}} = 10000$ days. With the log-uniform period prior, we keep the eccentricity prior uniform between 0 and 1. With the K18 prior, we test two eccentricity distributions: (1) $e$ uniform between 0 and 1 and (2) $e$ Beta-distributed, with parameters $a = 0.867, b=3.03$ adopted from \cite{kipping13}.

The results of these three sets of $P$ and $e$ prior choices for single transiter EPIC 246445793b (HIP 116454b) are plotted in Figure~\ref{fig:prior_comparison}. The K18 prior, as expected, yields a $P$ posterior which drops off much more steeply with increasing period than the log-uniform $P$ prior, regardless of the choice of $e$ prior.

In the case of EPIC 246445793b, the resulting K18 posteriors drop off so sharply with $P$ as to be $2\sigma$ inconsistent with the period measured from MOST photometry of $9.125\pm0.0005$ days \citep{vanderburg15}. We find that the K18 period posteriors for EPIC 211311380f (HIP 41378f) are similarly inconsistent with previously published period constraints \citep{becker19}.

These inconsistencies suggest that the K2 long-cadence transit data are insufficient to constrain our 12-parameter transit model in the face of such a strong period prior. It is possible that a lower-dimensional model---e.g., one that relied on a less flexible detrending algorithm than Gaussian process regression---would be less easily overwhelmed by a strong prior. It is also possible that the choice of intrinsic period prior power-law index $\alpha=-2/3$ is too steep to describe the long-period exoplanet population, but determining the intrinsic period prior for long-period exoplanets is well beyond the scope of this work.

As a result of these considerations and its success in the validation fits, we choose to fit the twelve single-transiters in our sample with a log-uniform prior in $P$ between $P_{\mathrm{min}}$ (determined by Equation~\ref{eq:Pmin}) and $P_{\mathrm{max}} = 10000$ days.

\subsection{Comparison to previous work}

For some of the twelve single transiters, period constraints have been published before. \cite{vanderburg15}, for example, obtained MOST photometry and radial velocity measurements of EPIC 246445793b and measured its period at $P = 9.1205 \pm 0.0005$ days. Our measurement, though obviously considerably less precise, agrees well with this one.

\cite{santerne18} analyze the K2 photometry of the EPIC
228801451 system and conclude that there are two possible orbital solutions for EPIC 228801451d: one with $P = 31.0 \pm 1.1$ days, and one with $P > 50$ days. Our period measurement, $P = 25.0^{+49.8}_{-0.1}$ days, is compatible with both.

\cite{vanderburg18} estimate the period of EPIC 248045685b by a very similar method to ours, also exploiting information from the host star's Gaia parallax and public broadband photometry. They conclude that $P = 106^{+74}_{-26}$ days, and we agree, with $P = 118^{+119}_{-41}$ days.

\cite{giles18} use high-resolution spectra combined with Gaia parallax to measure the stellar parameters of EPIC 248847494, and then model the transit of EPIC 248847494b to conclude that $P = 3650^{+1280}_{-1130}$ days, which agrees with our $P = 1700^{+4200}_{-400}$ days.

\cite{crossfield18} estimate that EPIC 212813907b has $P$ of order 1000 days, while we estimate that it has $P = 2000^{+1100}_{-300}$ days.

\cite{vanderburg16}, who discovered the EPIC 211311380 (HIP 41378) system in K2 Campaign 5 observations, estimate from fits to the respective single transits that planet d has period $P = 156^{+163}_{-78}$ days, planet e has period $P = 131^{+61}_{-36}$ days, and planet f has period $P = 324^{+121}_{-127}$ days. All three are consistent with our respective estimates of $49.3^{+328.5}_{-0.1}$ days, $131^{+250}_{-12}$ days, and $600^{+530}_{-160}$ days. \cite{becker19} observed subsequent transits of planets d and f in K2 Campaign 18 and refine the possible periods of all three planets based on the additional observations and simulations of dynamical stability. Their maximum-probability period for planet d is consistent with our $1\sigma$ credibility band, but their maximum-probability period for planet f (361 days) is shorter than we estimate. Our modal period for planet e is comfortably allowed by their dynamical simulations.

Finally, \cite{osborn16} fit their single transiters by an analogous method to ours; they derive effective temperatures for the planet host stars using broad-band photometry, then estimate stellar masses and radii from these temperatures using stellar models, assuming they are on the main sequence. Four single transits in our sample---EPIC 201635132b, EPIC 201892470b, EPIC 203311200b, and EPIC 204634789b---are drawn from their work.

However, for these four single transits, we do not agree with period determinations of \cite{osborn16}. This is because our stellar properties do not agree; in particular, the bulk densities we derive from Gaia parallaxes plus photometry do not agree with the bulk stellar densities calculable from their stellar mass and radius estimates. Only in the case of EPIC 203311200b are the period measurements even in $1\sigma$ agreement; for the other three, they are very different. Obtaining high-resolution spectra for these host stars would enable a precise cross-check on their stellar parameters and hopefully clear up this disagreement. (In the case of EPIC 204634789b, our $P_{\mathrm{min}}$ prior excludes their best-fit $P$, so consistent prior choices are also important.)

Finally, we note that four of our nine single transit fits---EPIC 201892470b, EPIC 204634789b, EPIC 228801451d (K2-229d), and EPIC 248045685b---have high modal posterior eccentricity, albeit with large uncertainty (see Table~\ref{tab:singles}). We are unable to achieve good fits to these transits with $e$ fixed to zero. These candidates, particularly EPIC 228801451d, which has two inner planetary companions, merit further study to determine their orbital properties more precisely.

\section{Conclusions}

In this work, we have presented new transit fits to twelve K2 single transiters, based on stellar density priors derived from GAIA parallaxes and publicly available broadband photometry. We achieve good precision in our period posteriors---when we let $e$ vary, the fractional $P$ uncertainty over the twelve single transiters is $94^{+87}_{-58}\%$, and when we fix $e=0$, it is $15^{+30}_{-6}\%$ (a roughly threefold improvement over typical period uncertainties of previous studies). In future, the best way to handle the question of eccentricity is likely to perform \texttt{single} transit fits with both fixed and free $e$, then use Bayesian model averaging to combine them and obtain a posterior $P$ estimate, but we leave this for future work.

Our fit period values also agree well with previously published period constraints. (Where we do not agree with earlier period constraints, it is because our stellar properties disagree with those used in previous work, e.g. \citealt{osborn16}; further study of these host stars, ideally with high-resolution spectroscopy, will be necessary to resolve the disagreement).

Additionally, we test this fitting method on 27 validation planets observed by K2. These planets have been observed to transit more than once, so their periods are known precisely; however, we model each transit individually, to evaluate the accuracy and precision of our derived $P$ posteriors given this limited information. We conclude that our method is robust as long as the individual transits we fit are well-sampled during ingress and egress, because our ability to measure $a/R_*$ from the transit shape is the most important predictive factor in the success of the method, both in terms of accuracy and precision.

TESS, with its relatively short 27.4 day observational baseline over much of the sky, is predicted to reveal tens to hundreds of new single transiters, and it will be necessary to estimate the periods of these planets to constrain their orbits, temperatures, surface and atmospheric properties, and potential habitability. Combining information from multiple surveys, including Gaia, promises to aid greatly in their characterisation.  

\section*{acknowledgements}
E.S.\ thanks David Kipping, Marcel Ag\"{u}eros, and the Columbia University President's Global Innovation Fund for their support, and Zephyr Penoyre for useful discussions and sharing his split normal fitting code.
N.E.\ would like to thank the Gruber Foundation for its generous support.
R.B.\ acknowledges support from FONDECYT Post-doctoral Fellowship Project No. 3180246., and from the Millennium Institute of Astrophysics (MAS).
A.J.\ acknowledges support from FONDECYT project 1171208 and by the Ministry for the Economy, Development, and Tourism's Programa Iniciativa Cient\'{i}fica Milenio through grant IC\,120009, awarded to the Millennium Institute of Astrophysics (MAS).


\bibliographystyle{mnras}
\bibliography{bib}

\bsp
\label{lastpage}

\end{document}